\documentstyle[aps,tighten,floats,psfig]{revtex}
\newcommand{\pp}{\bar{p}p}
\newcommand{\as}{\alpha_s}
\newcommand{\ra}{\rightarrow}
\newcommand{\papageno}{{\footnotesize PAPAGENO}}
\newcommand{\njets}{{\footnotesize NJETS}}
\newcommand{\cms}{{CMS}}
\newcommand{\mrs}{{\footnotesize MRS}}
\newcommand{\ehlq}{{\footnotesize EHLQ}}
\newcommand{\bcdms}{{\footnotesize BCDMS}}
\newcommand{\herwig}{{\footnotesize HERWIG}}
\newcommand{\pythia}{{\footnotesize PYTHIA}}
\newcommand{\isajet}{{\footnotesize ISAJET}}
\newcommand{\HERWIG}{{\footnotesize HERWIG 5.8}}
\newcommand{\PYTHIA}{{\footnotesize PYTHIA 5.6}}
\newcommand{\ISAJET}{{\footnotesize ISAJET 7.13}}
\newcommand{\chibz}{$\chi_{\footnotesize BZ}$}
\begin{document}

\date{\today}
\title{Studies of Topological Distributions of Inclusive Three-- and Four--Jet
      Events in $\pp$ Collisions at $\sqrt{s}=1800$~GeV with the D\O\ Detector}

% LIST_OF_AUTHORS.TEX                 07/31/95
%
\author{
%% names begin here
S.~Abachi,$^{12}$
B.~Abbott,$^{34}$
M.~Abolins,$^{23}$
B.S.~Acharya,$^{42}$
I.~Adam,$^{10}$
D.L.~Adams,$^{35}$
M.~Adams,$^{15}$
S.~Ahn,$^{12}$
H.~Aihara,$^{20}$
J.~Alitti,$^{38}$
% ...
\\
\vskip 0.50cm
\centerline{(D\O\ Collaboration)}
\vskip 0.50cm
}
\address{
\centerline{$^{1}$Universidad de los Andes, Bogot\'{a}, Colombia}
\centerline{$^{2}$University of Arizona, Tucson, Arizona 85721}
\centerline{$^{3}$Brookhaven National Laboratory, Upton, New York 11973}
\centerline{$^{4}$Brown University, Providence, Rhode Island 02912}
\centerline{$^{5}$University of California, Davis, California 95616}
\centerline{$^{6}$University of California, Irvine, California 92717}
\centerline{$^{7}$University of California, Riverside, California 92521}
\centerline{$^{8}$LAFEX, Centro Brasileiro de Pesquisas F{\'\i}sicas,
                  Rio de Janeiro, Brazil}
\centerline{$^{9}$CINVESTAV, Mexico City, Mexico}
\centerline{$^{10}$Columbia University, New York, New York 10027}
\centerline{$^{11}$Delhi University, Delhi, India 110007}
\centerline{$^{12}$Fermi National Accelerator Laboratory, Batavia,
                   Illinois 60510}
\centerline{$^{13}$Florida State University, Tallahassee, Florida 32306}
\centerline{$^{14}$University of Hawaii, Honolulu, Hawaii 96822}
\centerline{$^{15}$University of Illinois at Chicago, Chicago, Illinois 60607}
\centerline{$^{16}$Indiana University, Bloomington, Indiana 47405}
\centerline{$^{17}$Iowa State University, Ames, Iowa 50011}
\centerline{$^{18}$Korea University, Seoul, Korea}
\centerline{$^{19}$Kyungsung University, Pusan, Korea}
\centerline{$^{20}$Lawrence Berkeley Laboratory and University of California,
                   Berkeley, California 94720}
\centerline{$^{21}$University of Maryland, College Park, Maryland 20742}
\centerline{$^{22}$University of Michigan, Ann Arbor, Michigan 48109}
\centerline{$^{23}$Michigan State University, East Lansing, Michigan 48824}
\centerline{$^{24}$Moscow State University, Moscow, Russia}
\centerline{$^{25}$University of Nebraska, Lincoln, Nebraska 68588}
\centerline{$^{26}$New York University, New York, New York 10003}
\centerline{$^{27}$Northeastern University, Boston, Massachusetts 02115}
\centerline{$^{28}$Northern Illinois University, DeKalb, Illinois 60115}
\centerline{$^{29}$Northwestern University, Evanston, Illinois 60208}
\centerline{$^{30}$University of Notre Dame, Notre Dame, Indiana 46556}
\centerline{$^{31}$University of Oklahoma, Norman, Oklahoma 73019}
\centerline{$^{32}$University of Panjab, Chandigarh 16-00-14, India}
\centerline{$^{33}$Institute for High Energy Physics, 142-284 Protvino, Russia}
\centerline{$^{34}$Purdue University, West Lafayette, Indiana 47907}
\centerline{$^{35}$Rice University, Houston, Texas 77251}
\centerline{$^{36}$Universidade Estadual do Rio de Janeiro, Brazil}
\centerline{$^{37}$University of Rochester, Rochester, New York 14627}
\centerline{$^{38}$CEA, DAPNIA/Service de Physique des Particules, CE-SACLAY,
                   France}
\centerline{$^{39}$Seoul National University, Seoul, Korea}
\centerline{$^{40}$State University of New York, Stony Brook, New York 11794}
\centerline{$^{41}$SSC Laboratory, Dallas, Texas 75237}
\centerline{$^{42}$Tata Institute of Fundamental Research,
                   Colaba, Bombay 400005, India}
\centerline{$^{43}$University of Texas, Arlington, Texas 76019}
\centerline{$^{44}$Texas A\&M University, College Station, Texas 77843}
}
%end

\maketitle

\vspace*{1.0cm}
\centerline{\large Abstract}

\begin{abstract} The global topologies of inclusive three-- and four--jet
events produced in $\pp$ interactions are described. The three-- and four--jet
events are selected from data recorded by the D\O\ detector at the Tevatron
Collider operating at a center--of--mass energy of $\sqrt{s} = 1800$ GeV. The
measured, normalized distributions of various topological variables are
compared with  parton--level predictions of tree--level QCD calculations.  The
parton--level QCD calculations are found to be in good agreement with the
data. The studies also show that the topological distributions of the different
subprocesses involving different numbers of quarks are very similar and
reproduce the measured distributions well. The parton shower Monte Carlo
generators provide a less satisfactory description of the topologies of the
three-- and four--jet events.
\end{abstract}

\section*{Introduction}

The Fermilab Tevatron Collider provides a unique opportunity to study  the
properties of strong interactions in $\pp$ collisions at short distances.  The
hard scattering is described by the theory of perturbative Quantum
Chromodynamics (QCD)~\cite{QCD1,QCD2,QCD3} and has been studied extensively  in
the last decade~\cite{QCDEE,QCDPP}.  Within the context of QCD, the hard
process is described as a point--like  scattering between constituent partons
(quarks and gluons) of protons and  anti--protons. The scattering cross
sections can be written in  expansions in terms of powers of the strong
coupling constant $\alpha_s$ convoluted with  parton momentum distributions
inside the nucleon.  The lowest order  $\alpha_s^2$ term corresponds to the
production of two--parton final states. Terms of order $\as^3$ and $\as^4$ in
the expansion imply the existence of  three-- and four--parton final states,
respectively. Colored partons from the hard scattering evolve via soft quark
and gluon  radiation and hadronization processes to form observable colorless
hadrons,  which appear in the detector as localized energy deposits identified
as jets. High energy jets originating from partons in the initial hard
scattering process are typically isolated from other collision products. They
are expected to preserve the energy and direction of the  initial partons, and
therefore the topologies of the final jet system are assumed to be directly
related to those of the initial parton system.

The cross section and angular distributions for two--jet events have been
successfully compared with the predictions of QCD~\cite{QCDPP,2JET}.  A study
of three-- and four--jet events allows a test of the validity  of the QCD
calculations to higher order ($\as^3$ or beyond) and a probe of the underlying
QCD dynamics. This paper explores the topological distributions of three-- and
four--jet events. The distributions provide sensitive tests of the QCD matrix
element calculations.  Topological distributions for the three-- and four--jet
events have been published previously by the UA1, UA2 and CDF
Collaborations~\cite{3JET,CDF3,UA24,CDF4}. However,  all of these studies
imposed requirements on the topological variables themselves, and therefore
significantly reduced the phase space under study. This paper extends these
studies to previously untested regions of phase space for a large number of
topological variables.  The measured normalized distributions, without
restrictions on the topological variables themselves, are compared with the QCD
tree--level matrix element calculations.  The predictions from simple
phase--space matrix elements are shown as a comparison, and the distributions
of QCD subprocesses involving different numbers of  quarks are also examined.
Finally, the data are compared with the predictions of  three parton shower
event generators.

\section*{Definition of Topological Variables}

The topological variables used in this paper are defined in the parton or
jet center--of--mass system (\cms). The definitions refer to partons and
jets interchangeably. The partons are assumed to be massless and the
jet masses are ignored by using the measured jet energies as the magnitudes
of jet momenta.

\begin{figure}[htb]
  \begin{center}
    \begin{tabular}{ll}
      \psfig{figure=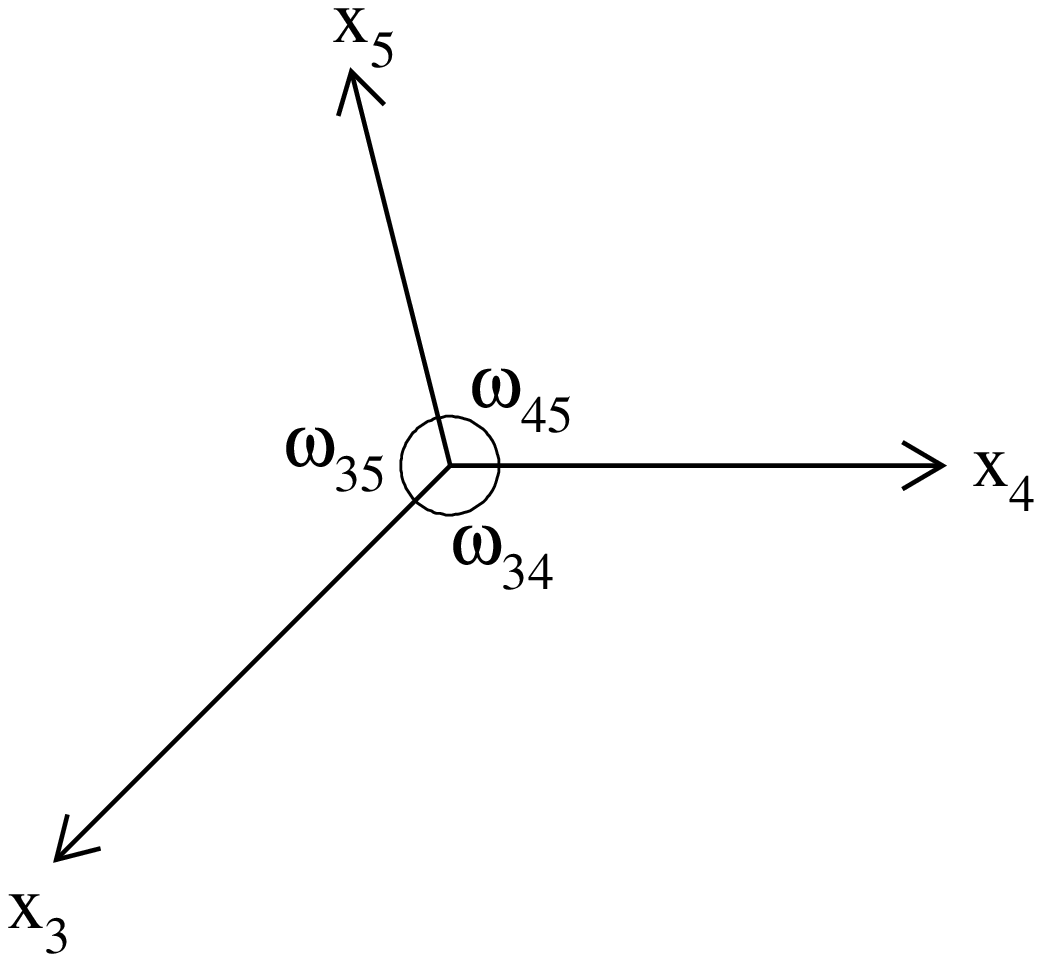,width=9.0cm} &
      \psfig{figure=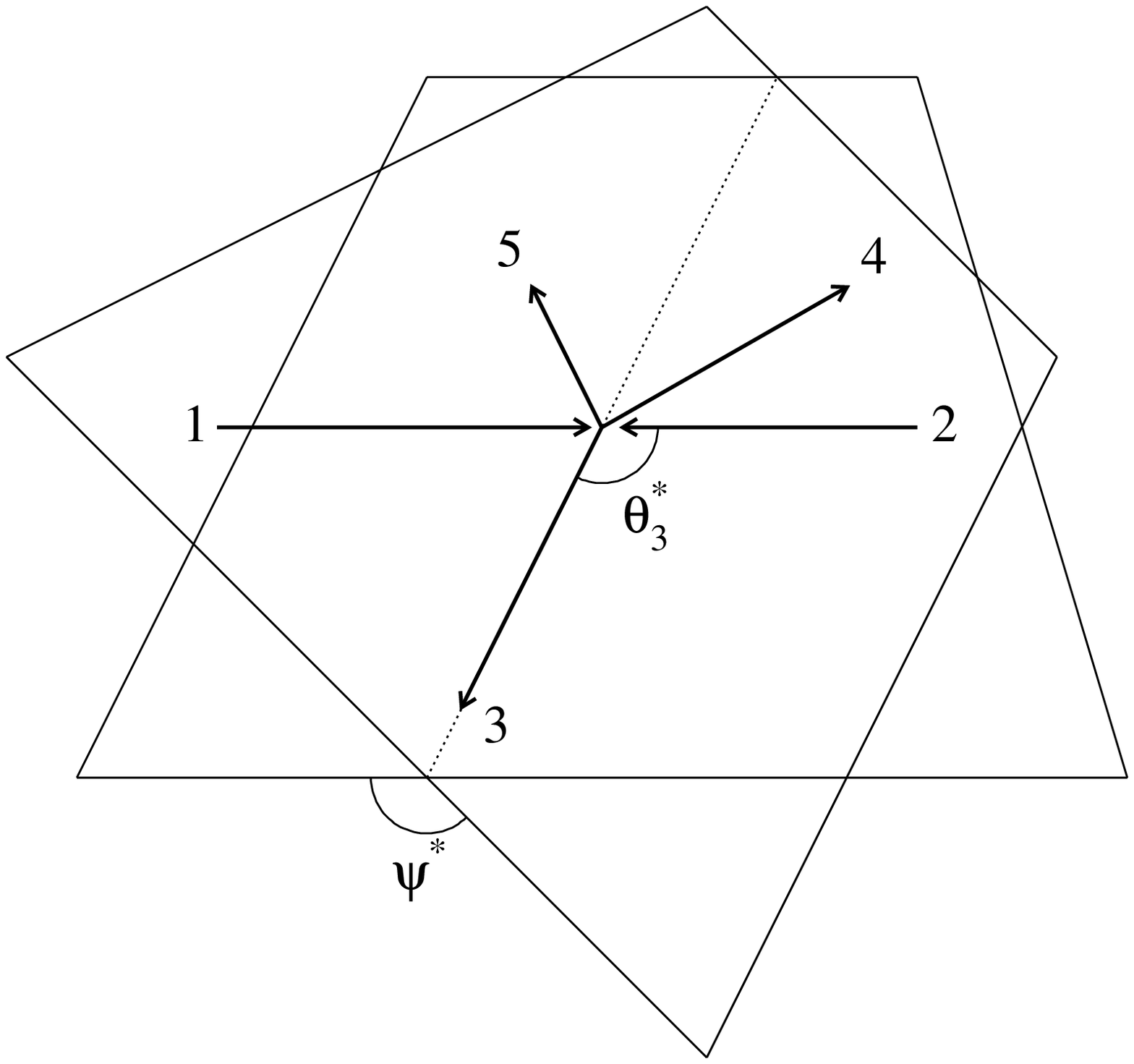,width=8.0cm} \\
    \end{tabular} \\
    \caption{Illustration of the topological variables:
             $x_i$, $\omega_{ij}$, $\theta^*_3$ and $\psi^*$
             for the three--jet events.}
    \label{fig:3jfig}

    \begin{tabular}{ll}
      \psfig{figure=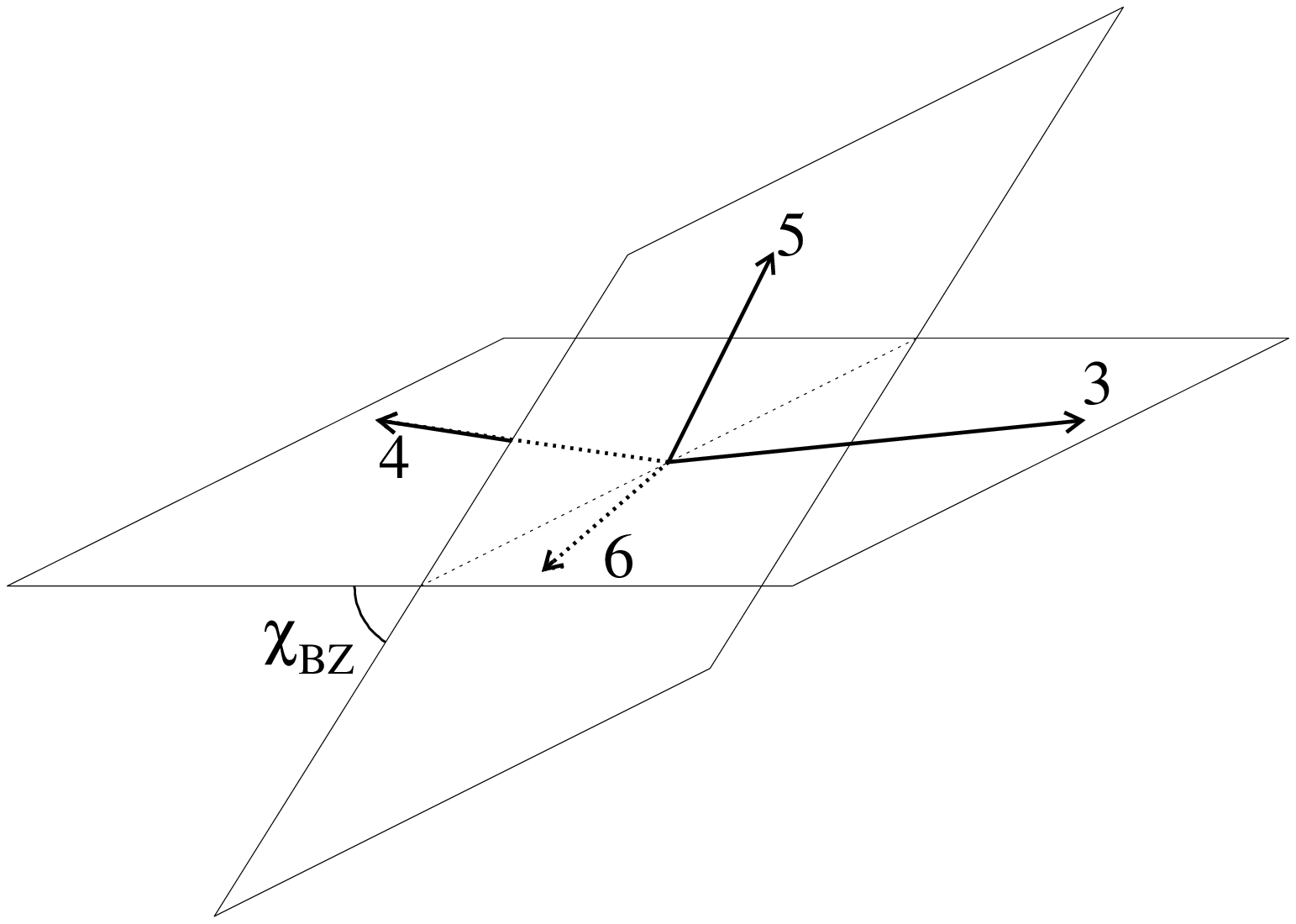,width=9.0cm} &
      \psfig{figure=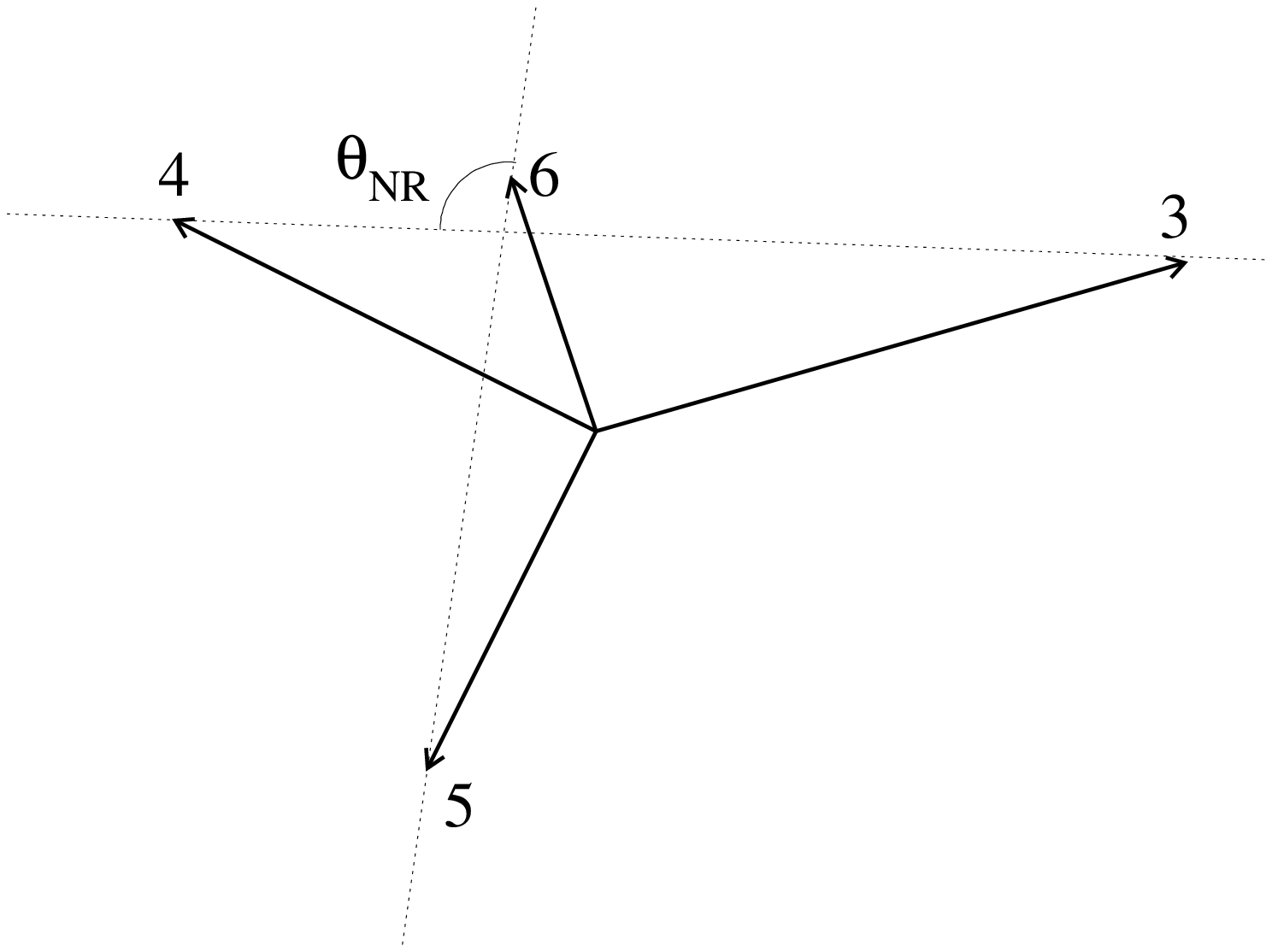,width=9.0cm} \\
    \end{tabular} \\
    \caption{Illustration of $\chi_{BZ}$ and $\theta_{NR}$ definitions for
             the four--jet events.}
    \label{fig:4jfig}
  \end{center}
\end{figure}

The topological properties of the three--parton final state in the
center--of--mass system  can be described in terms of six variables. Three of
the variables reflect partition of the \cms\ energy among the three
final--state partons. The other three variables  define the spatial orientation
of the planes containing the three partons. It is convenient to introduce the
notation $1+2\rightarrow 3+4+5$ for  the three--parton process. Here, numbers 1
and 2 refer to incoming partons while the numbers 3, 4 and 5 label the outgoing
partons, ordered in descending \cms\  energies, i.e. $E_3 > E_4 > E_5$. The
final state parton energy is an  obvious choice for the topological variables
for the three--parton final state. For simplicity,  $E_i\ (i=3,4,5)$ is often
replaced by the scaled variable $x_i\ (i=3,4,5)$, which is defined by  $x_i =
2E_i/\sqrt{\hat s}$, where $\sqrt{\hat s}$ is the center--of--mass energy of
the hard scattering process. By definition, $x_3 + x_4 + x_5 = 2$. The scaled
parton energies $x_i$  and the angles between partons ($\omega_{jk},\
j,k=3,4,5$) for  the three--parton final state have the following relationship:
\begin{equation} x_i = \frac{2\sin\omega_{jk}}
{\sin\omega_{34}+\sin\omega_{45}+\sin\omega_{53}}, \end{equation} where
$i,j,k=3,4,5$ and $i\not=j\not=k$.  Clearly, the internal structure of the
three--parton final state is completely determined by any two scaled parton
energies. The angles that fix the event orientation can be  chosen to be: (1)
the cosine\footnote{Unless  otherwise specified, the absolute values of the
cosines of polar angles  are implied throughout this paper.} of the polar angle
with respect to the beam ($\cos\theta^*_3$) of parton 3, (2) the azimuthal
angle of parton 3 ($\phi^*_3$), and (3) the angle between the plane containing
partons 1 and 3 and the plane containing partons 4 and 5 ($\psi^*$) defined by:
\begin{equation} \cos\psi^* =
\frac{(\vec{p}_1\times\vec{p}_3)\cdot(\vec{p}_4\times\vec{p}_5)}
{|\vec{p}_1\times\vec{p}_3| |\vec{p}_4\times\vec{p}_5|},  \end{equation} where
$\vec{p}_i$ is the parton momentum. Figure~\ref{fig:3jfig} illustrates the
definition of the topological variables for the three--parton final state. For
unpolarized beams (as at the Tevatron), the $\phi^*_3$ distribution  is
uniform. Therefore, only four independent kinematic variables are needed  to
describe the topological properties of the three--parton final state. In this
paper,  they are chosen to be $x_3$, $x_5$, $\cos\theta_3^*$ and $\psi^*$.

Another set of interesting  variables is the scaled invariant mass of jet
pairs: \begin{equation} \mu_{ij} = \frac{m_{ij}}{\sqrt{\hat s}}\equiv\sqrt{x_i
x_j (1-\cos\omega_{ij})/2} \ \ i,j=3,4,5\ \ {\rm and}\ i\not=j, \end{equation}
where $m_{ij}$ is the invariant mass of partons $i$ and $j$ and $\omega_{ij}$
is the opening angle between the two partons. The scaled invariant mass
($\mu_{ij}$)  is sensitive to the scaled energies of the two partons, the angle
between  the two partons, and the correlations between these variables. Using
dimensionless variables and making comparisons of normalized distributions
minimizes the systematic errors due to detector resolution  and jet energy
scale uncertainty and therefore allows a direct comparison  between data and
theoretical calculation.

The four--parton final state is more complicated. Apart  from the \cms\ energy,
eight independent parameters are needed to completely  define a four--parton
final state in its center--of--mass system. Two of these define the overall
event orientation while the other six fix the  internal structure of the
four--parton system.  In contrast to the three--parton final state, there is no
simple relationship between the scaled parton energies and the opening angles
between partons. Consequently, the choice of  topological variables is less
obvious in this case. In this paper,  variables are defined in a way similar to
those investigated for the three--parton final state. The four partons are
ordered in descending \cms\ energy and labeled from 3 to 6. The variables
include the scaled energies ($x_i,\ i=3,...,6$), the cosines of polar angles
($\cos\theta^*_i,\ i =3,...,6$) of the four jets, the cosines of the opening
angles between partons  ($\cos\omega_{ij},\ i,j=3,...,6$, and $i\not=j$), and
the scaled masses ($\mu_{ij} = m_{ij}/\sqrt{\hat s},\ \ i,j=3,...,6$ and
$i\not=j$) of parton pairs. In addition, two variables characterizing the
orientation of event planes are investigated. One of the two variables is the
``Bengtsson--Zerwas'' angle (\chibz)~\cite{BZ} defined as the angle between the
plane containing the two leading jets and the plane containing the two
non--leading jets: \begin{equation} \cos\chi_{BZ} = \frac{(\vec{p}_3\times
\vec{p}_4)\cdot (\vec{p}_5\times \vec{p}_6)} {|\vec{p}_3\times \vec{p}_4|
|\vec{p}_5\times \vec{p}_6|}. \end{equation} The other variable is the cosine
of the ``Nachtmann--Reiter'' angle ($\cos\theta_{NR}$)~\cite{NR} defined as the
angle between the momentum vector differences of the two leading jets and the
two non--leading jets: \begin{equation} \cos\theta_{NR} =
\frac{(\vec{p}_3-\vec{p}_4)\cdot (\vec{p}_5-\vec{p}_6)}
{|\vec{p}_3-\vec{p}_4| |\vec{p}_5-\vec{p}_6|}. \end{equation}
Figure~\ref{fig:4jfig} illustrates the
definitions of \chibz\ and $\theta_{NR}$ variables. Historically, $\chi_{BZ}$
and $\cos\theta_{NR}$ were proposed for $e^+e^-$  collisions to study gluon
self--coupling. Their interpretation in $\pp$ collisions is more complicated,
but the variables can be used as a tool for studying the internal structure of
the four--jet events.

\section*{The Theoretical Model}

The cross section for the production of the $n$--parton final state
$1+2\rightarrow 3+ \cdots + (n+2)$, in $\pp$ collisions at a center--of--mass
energy $\sqrt{s}$ is described by the following expression: \begin{equation}
\sigma_n = \sum_{\ell}\int f^\ell_1(x_1) f^\ell_2(x_2) |M^n_\ell|^2  \Phi_n
dx_1 dx_2, \end{equation} where the sum runs over all possible $1+2\rightarrow
n-$parton  subprocesses. The functions $f^\ell_1(x_1)$ and $f^\ell_2(x_2)$ are
the parton  density functions of the incoming partons,  $|M^n_\ell|^2$
represents the matrix elements of the subprocess, and $\Phi_n$ is the $n$--body
phase space.  Theoretically $|M^n_\ell|^2$ is well
behaved if calculated to all orders in the $\as$ expansion. At present, this
calculation is technically not possible and one has to deal with truncated
expansion. As a result, $|M^n_\ell|^2$ diverges when the energy of any final
state parton or the angle between any two partons approaches zero. The
singularities in $|M^n_\ell|^2$ cause poles in the  topological distributions.
In comparison, a phase--space model in which $|M^n_\ell|^2 \propto 1/{\hat
s}^{n-2}$ where ${\hat s} = x_1 x_2 s$  does not have singularities in the
matrix element, therefore, the  topology of the model is determined by the
phase space $\Phi_n$.  In this paper, the distributions from the phase--space
model are used as references for the comparisons between the data and QCD.

Presently two approaches for modeling perturbative QCD for multi--jet
production exist. The straight--forward method is the matrix element method, in
which Feynman diagrams are calculated order--by--order in $\as$. Technical
difficulties have limited the calculations to the tree--level of the relevant
processes.  The exact tree--level matrix element calculation for the
three--parton  final state has been available for some time~\cite{ME3}. The
complete  tree--level matrix element calculations for up to five final state
partons  have been recently calculated by Berends, Giele and
Kuijf~(BGK)~\cite{BGK} using  a Monte Carlo method. The other commonly used
approximate calculations are those of Kunszt and Stirling~(KS)~\cite{KS} and of
Maxwell~\cite{MAX}.  The perturbative QCD calculations have been incorporated
into several  partonic event generators.  The exact tree--level matrix elements
calculations  for up to five jets are implemented in the \njets~\cite{BGK}
program.  \papageno~\cite{PAPAGENO} implements an exact matrix element
calculation of tree--level contributions for final states with up to three
partons and provides KS and Maxwell approximations for up to six partons. These
approximations are used in part to  speed up the calculations, in view of the
complicated exact matrix elements.  For the analysis described in this paper,
the \njets program is used to calculate QCD predictions while the \papageno
program is used as a cross check and to calculate distributions from the
phase--space model.

The second approach is based on the parton shower scheme. In this method, the
hard scattering begins with two initial outgoing partons. An arbitrary number
of partons are then branched off from the two outgoing partons and the two
incoming partons (backward evolution) to yield a description for multi--parton
production, with no explicit upper limit on the number of partons involved. The
parton shower picture is derived within the framework of the Leading
Logarithmic Approximation (LLA)~\cite{LLA}. As a result of the approximation,
the reliability of the parton shower is expected to decrease as parton
multiplicity increases. Many parton shower Monte Carlo event generators are
available. In this paper, \HERWIG~\cite{HERWIG}, \ISAJET~\cite{ISAJET} and
\PYTHIA~\cite{PYTHIA} are used.

\section*{The Data Sample}

The data used in this analysis were collected with the D\O\ detector during the
1992--1993 Tevatron run at a center--of--mass energy of 1800 GeV. The D\O\
detector consists of a central tracking system, a calorimeter, and muon
chambers. Jets are measured in the calorimeter, which has a transverse
segmentation of $\Delta\eta\times\Delta\phi = 0.1\times 0.1$. The jet energy
resolution is typically 15\% at $E_T$=50~GeV  and 7\% at
$E_T$=150~GeV~\cite{JETRESO}.  The jet direction is measured with a resolution
of 0.05 in both $\eta$ and $\phi$. With the hermetic and uniform rapidity
coverage  ($-4.5 < \eta < 4.5$) of the calorimeter, the D\O\ detector is well
suited for studying multi--jet physics.  A detailed description of the D\O\
detector can be found elsewhere~\cite{DO}.

The events used in this study passed  hardware (Level~1) and   software
(Level~2) energy--cluster based triggers. In addition,  a Level~0 hardware
trigger required that vertices along the beam line be within  10.5~cm of $z=0$.
The Level~1 trigger was based  on energy deposited in calorimeter towers of
size  $\Delta\eta\times\Delta\phi = 0.2\times 0.2$. The events were required to
have at least two such towers with transverse energy ($E_T$) above 7~GeV. The
successful candidates were passed to the Level~2  trigger, which summed
transverse energies of calorimeter towers in a cone of radius ${\cal
R}(\equiv(\Delta\eta^2 + \Delta\phi^2)^{1/2})=0.7$. The Level~2 trigger
selected those events with at least one such cone, built around the Level~1
trigger tower,  with transverse energy above 50~GeV.  The total effective
luminosity used in this analysis is 1.2 pb$^{-1}$. The trigger efficiency for
events with at least one jet with $E_T > 60$~GeV is above 90\%~\cite{FSIMU}. A
detailed description of the trigger can be found elsewhere~\cite{TRIG}.

The offline reconstruction uses a fixed--cone jet algorithm  with ${\cal
R}=0.7$, similar to the algorithm used in the Level~2 trigger.  The jet
reconstruction begins with seed calorimeter towers of size
$\Delta\eta\times\Delta\phi=0.1\times 0.1$ containing more than 1 GeV
transverse energy. Towers are represented by massless  four--momentum vectors
with directions given by the tower positions and event vertices. The four
momenta of towers in the cone around the seed tower are summed to form the
four--momentum vector of the jet. The jet direction is then recalculated using
tower directions weighted by their transverse energies. The procedure is
repeated until the jet axis converges. For two overlapping jets, if either jet
shares more than 50\% of its transverse energy with the other jet, the two jets
are merged. Otherwise they are split and the shared transverse energy is
equally divided between  the two jets. The final jet $E_T$ is the sum of the
transverse energies of towers within the cone, while the jet direction is
determined by the jet four--momentum vector $(E,E_x,E_y,E_z)$, i.e.,
$\theta=\cos^{-1}(E_z/\sqrt{E_x^2+E_y^2+E_z^2})$, $\phi=\tan^{-1}(E_y/E_x)$ and
$\eta=-\ln\tan(\theta/2)$.

The jet energy scale has been calibrated using direct photon candidates by
balancing jet $E_T$ against that of the photon candidate. The electromagnetic
energy scale was determined by comparing the measured electron pair mass  of
$Z\ra e^+e^-$ events with the $Z$ mass~\cite{ZMASS} measured by  $e^+e^-$
experiments. The calibration takes into account the effects of out--of--cone
particle showering using shower profiles from test beam data as well as the
underlying event using events from minimum--bias triggers. Details can be found
in Ref.~\cite{CAL}.

After energy corrections, jets are required to have $E_T$ greater  than 20~GeV
and lie within  a pseudorapidity range of $-3.0$ to 3.0. The pseudorapidity is
calculated  with respect to the event vertex determined from tracks measured by
the central tracking detector. Jets passing the above criteria are ordered in
decreasing  $E_T$. The $E_T$ of the leading jet must be greater than 60~GeV to
reduce possible trigger bias and threshold effects.

\begin{figure}[htb]
  \begin{center}
      \mbox{\psfig{figure=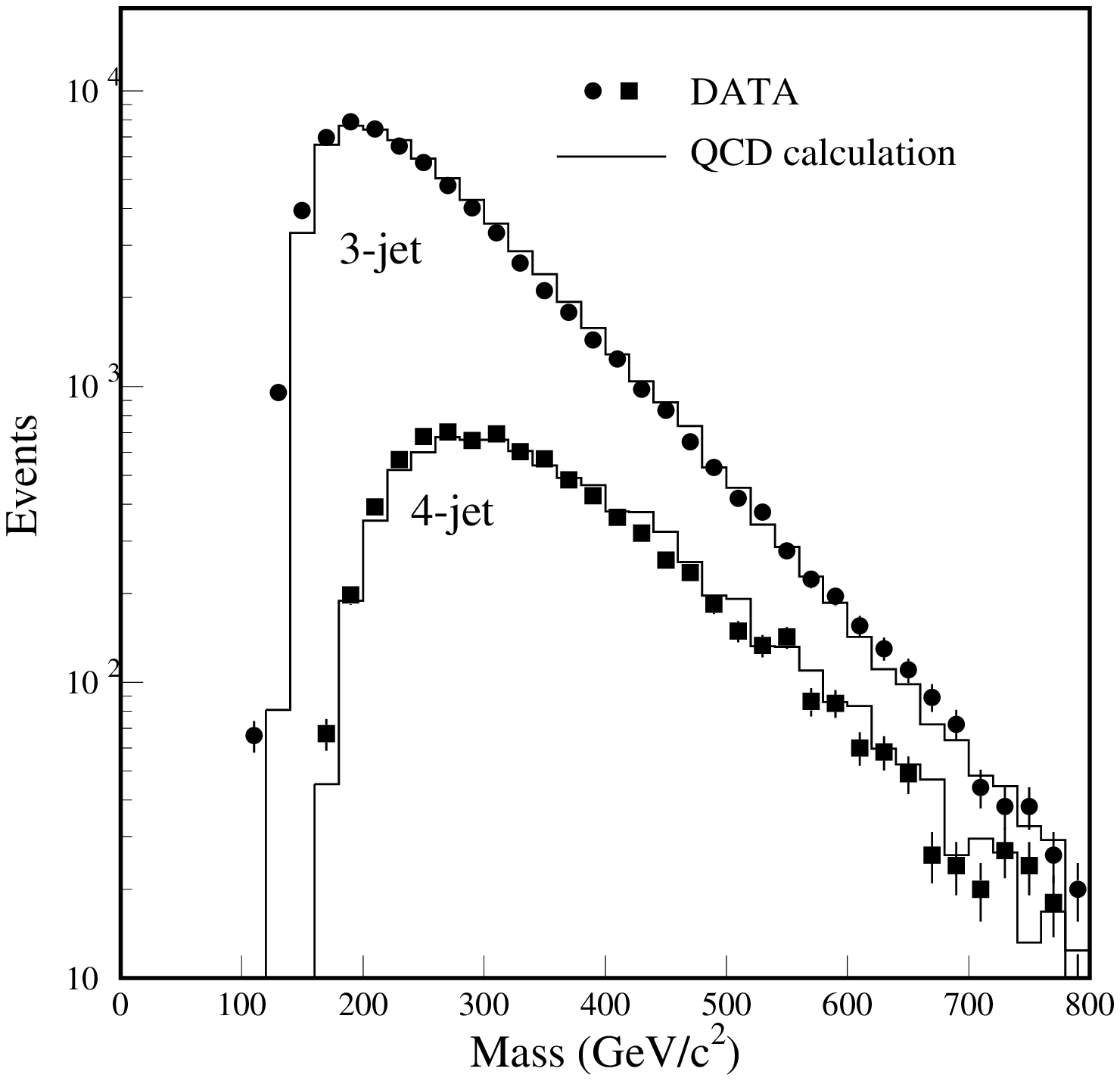,width=8.0cm}}
    \caption{The mass distributions of the selected three-- and four--jet
events
             before the mass requirement. The QCD matrix element calculations
             are normalized to the data.}
    \label{fig:3jmjjj}

\vspace*{1.0cm}
    \begin{tabular}{ll}
      \psfig{figure=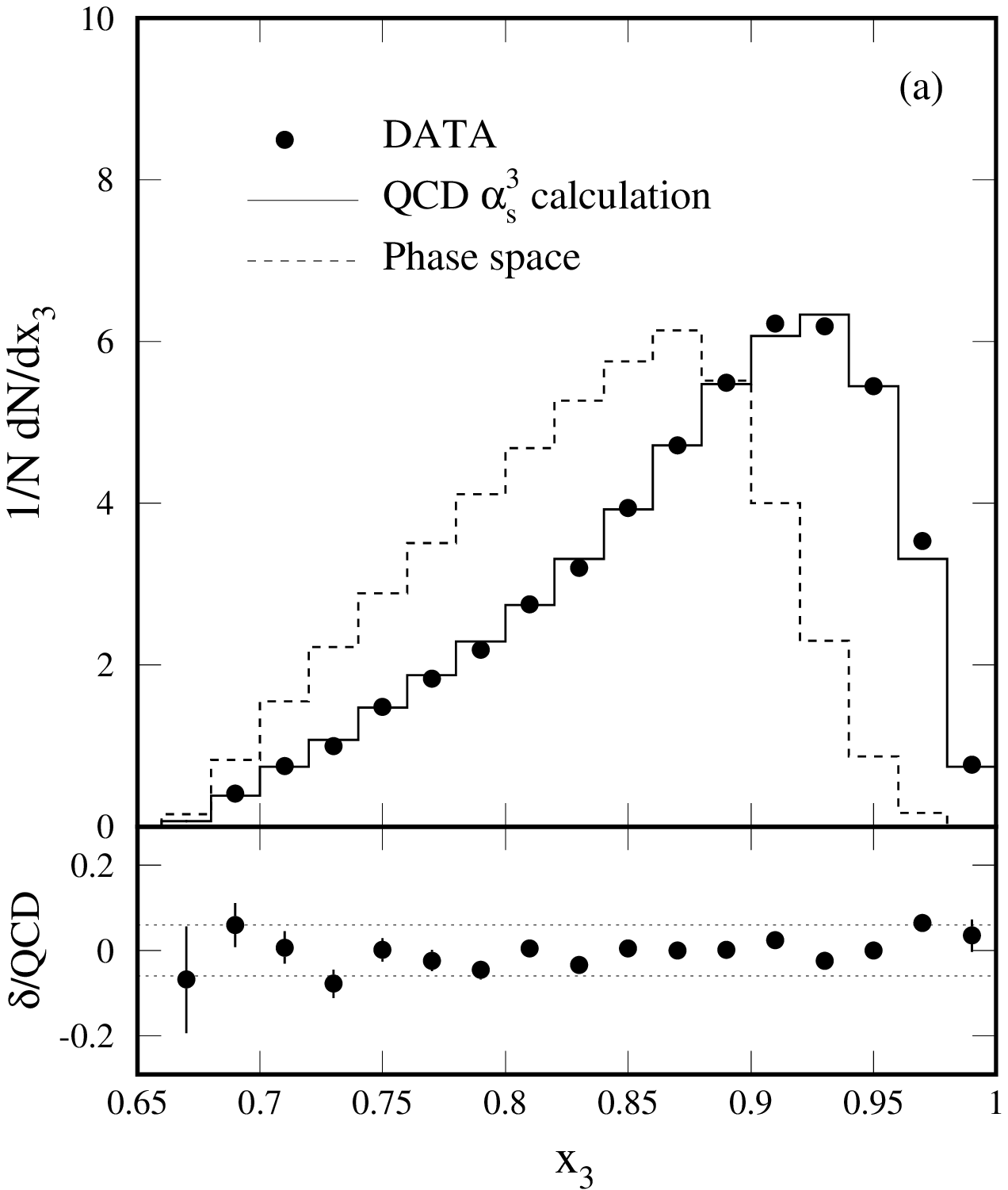,width=8.0cm} &
      \psfig{figure=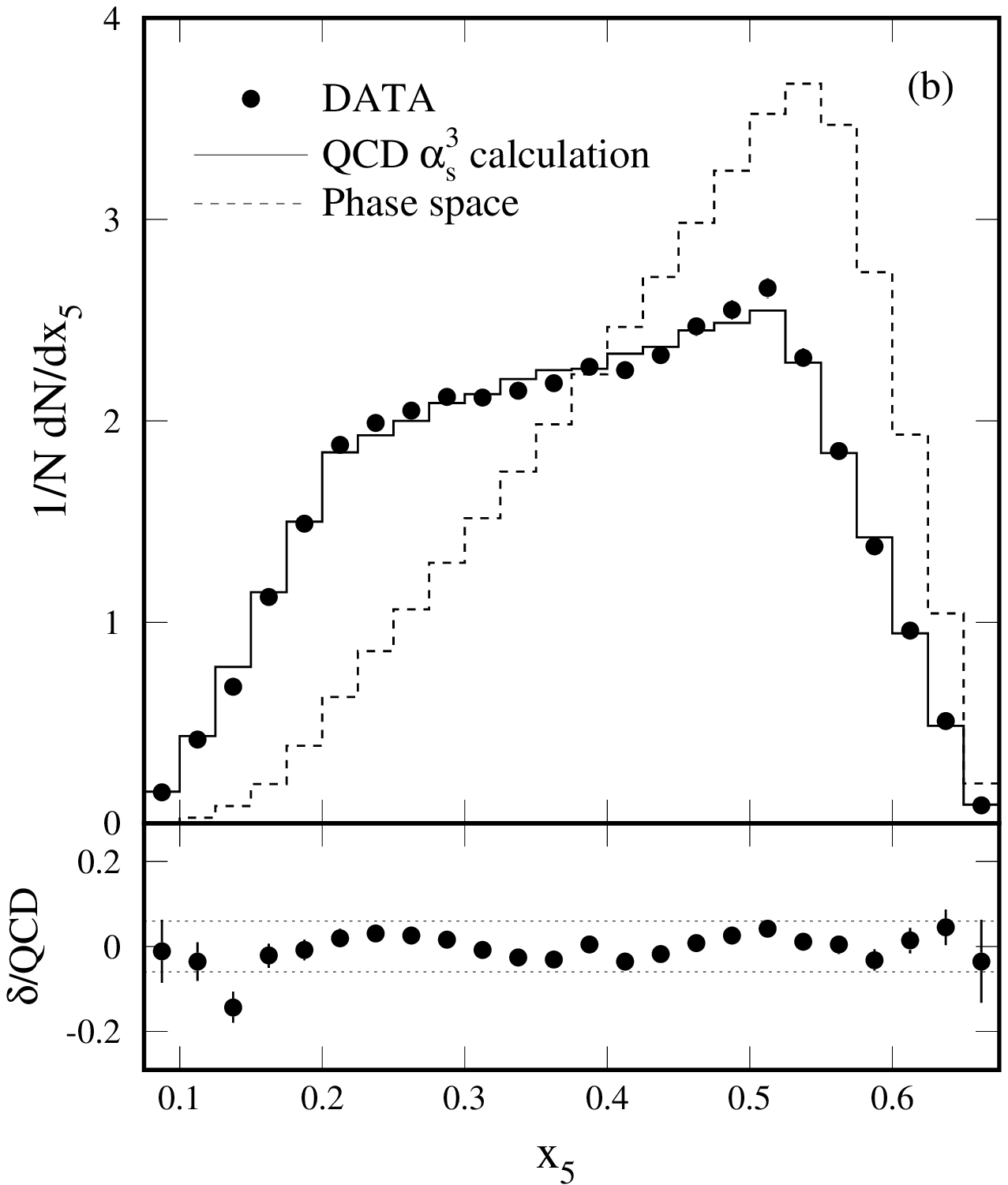,width=8.0cm} \\
    \end{tabular} \\
    \caption{The scaled energy distributions of (a) $x_3$ and (b) $x_5$
             for three--jet events in their center--of--mass system.
             Only statistical errors are shown. The bottom plot shows
             the fractional difference of the data from the exact
             tree--level QCD calculation. The RMS of the fractional differences
             is 3.4\% for $x_3$ and 3.9\% for $x_5$.
             The dotted lines show the
             estimated 6\% systematic uncertainty on the measurement.
             The data are also listed in Table~\protect\ref{tab:3jx35}.}
    \label{fig:3jx35}
  \end{center}
\end{figure}

Three--jet events are selected by further demanding that there be at least
three jets. This leaves about 94,000 events in the sample. The separation
$\Delta{\cal R}$ between jets is required to be greater than 1.4,  which is
twice the cone size used, to avoid systematic uncertainty associated with the
merging/splitting of the cone jet algorithm. This requirement removes events
with overlapping  jets and therefore ensures good jet energy and direction
measurements.  Approximately 70\% of the events pass this requirement. The
invariant mass distribution of the three highest $E_T$ jets is  shown in
Fig~\ref{fig:3jmjjj}. Also shown is the distribution from the exact tree--level
calculations of perturbative QCD. The overall agreement between the data and
QCD distributions is good with the exception of the low mass region, where the
threshold and resolution effects are important. To reduce possible bias in this
region, the invariant mass of the three leading jets is required to be above
200~GeV/c$^2$. After all selection criteria, a sample of about 46 thousands
three--jet events remains.  The surviving events are then transformed to the
\cms\ frame of the three  leading jets. Any other jets in the event are
ignored. The jets are re--ordered in descending energy in their \cms\ system.
The topological variables ($x_3$, $x_4$, $\cos\theta^*_3$ and $\psi^*$) are
calculated. Unlike previous studies by other experiments, no requirements on
these topological  variables are imposed. If the topological requirements
similar to those in Ref.~\cite{CDF3} were imposed, the three--jet event sample
would be reduced by more than a  factor of ten.

Four--jet events are selected in a similar manner. Events are required to have
at least four jets, which results in a data sample of 19,000 events. The
$\Delta{\cal R}$ between any jet pair  is required to be greater than 1.4,
reducing the data sample to about 8,400 events. As in the selection of the
three--jet events, the invariant mass of  the leading four jets must be above
200~GeV/c$^2$. The mass distribution before this requirement is applied is
shown in Fig.~\ref{fig:3jmjjj}. A total of 8,100 events remains in the
four--jet event sample.  The four leading jets of the remaining events are
boosted to their center--of--mass  system, and are ordered in decreasing
energy.  Additional jets, if present, are ignored.  The topological variables
are calculated using the four boosted momentum vectors after ordering in
decreasing energy. No requirements on the topological variables are imposed.

\section*{Predictions of Theoretical Models}

The partonic event generator \njets\ is used to calculate the exact
tree--level QCD distributions. The \papageno\ program is used to calculate the
distributions of the phase--space model as well as  the approximate
calculations of KS. Unless otherwise specified, the parton distribution
function used in the calculations is \mrs~(\bcdms\ fit)~\cite{MRS} for both
\njets\ and \papageno.  The QCD scale parameter is set to 200 MeV and the
renormalization scales are set to the average transverse momentum  of the
outgoing partons for both matrix elements and parton distribution functions.
The outgoing partons are analyzed as if they were observed  jets and the
selection criteria described above are applied to select three--  and four--jet
events.

To study the sensitivity to the choice of parton distribution function, the
topological distributions of QCD calculations with different parton
distribution  functions are compared. For \njets, the comparisons are made
between  \mrs~\cite{MRS} and \ehlq~\cite{EHLQ} parton distribution functions.
For \papageno\ the parton distribution functions of \mrs~\cite{MRS} and
Morfin--Tung~\cite{SFMT} are employed. Although the total three-- and four--jet
cross sections vary by as much as 30\% for different parton distribution
functions, the normalized  topological distributions are found to be very
insensitive to the choice.  A typical difference of less than 3\% is found for
the variables studied. The dependences on the renormalization scale are
investigated using the \papageno\ program. The distributions for the
renormalization scales of (1) the average transverse momentum, (2) one half the
average value of transverse momentum and (3) the total transverse energy are
compared.  Despite large differences (as much as 60\%) in the total production
cross sections, the differences between normalized distributions are very
small, typically less than 3\%. Combining the effects described above,  the
uncertainty on the theoretical predictions is 4\%.

The fragmentation effect is investigated using the \herwig\ event generator.
The parton--level distributions for three-- and four--jet events are compared
with the distributions at particle level. For parton shower Monte Carlo
programs, partons are defined  as those quarks and gluons after the parton
showering and before the fragmentation. The differences between the
distributions before and after fragmentation are found to be small, typically
at less than 4\% level. The small fragmentation effect combined with a small
detector effect discussed below enable direct comparisons between data and
theoretical parton--level  calculations.

Both \njets\ and \papageno\ incorporate tree--level calculations for three--
and four--parton final states. The effect on the normalized distributions due
to higher--order loop corrections is expected~\cite{GIELE} to be small in the
phase--space region relevant to the analyses described in this  paper. Although
both Monte Carlo programs generate exclusive events, the three or four jets of
the generated events predict the behavior  of the leading three or four jets of
an inclusive data sample~\cite{GIELE}. Therefore the data distributions based
on the inclusive samples are compared with QCD calculations from exclusive
final states in this paper.

\section*{Uncertainties of the Measured Topological Distributions}

The measured distributions of topological variables are affected by: (a) the
trigger efficiency, (b) the detector  acceptance and resolution, and (c) the
uncertainty of the energy scale.  However, most of these corrections and their
uncertainties are  minimized by normalizing the distributions to unit area and
by selection requirements.  In the following, residual uncertainties are
discussed.

The non--uniformity of the detector acceptance and of the trigger efficiency in
the topological variables and the detector energy resolution and angular
resolution have direct effects on the measured distributions. These effects are
estimated using a fast detector simulation  program~\cite{FSIMU} which takes
into account the detector energy and angular resolution and the trigger
efficiency as functions of the pseudorapidity and the  transverse energy of
jets. The bin--by--bin correction factors are flat within  5\%.

By definition, the topological variables have a weak dependence on the energy
scale since only the scaled energies and directions of  the jets are used.
However, the event selection criteria, such as $E_T$ and invariant mass
requirements, are vulnerable to the energy scale error. The possible distortion
of the measured topological variables due to  the uncertainty in the energy
scale is studied by varying the energy  calibration constants within their
nominal errors. The selection procedure described above is repeated for the
events  calibrated with these modified constants. Apart from some low
statistics  bins, the variations in the measured topological variables are very
small.  We conservatively assign a 3\% systematic error on the topological
distributions due to energy scale uncertainty. The small variation is  in part
due to the fact that the topological distributions change slowly with the jet
$E_T$ and the invariant mass of the jet system.

In principle, the measured distributions have to be corrected for detector
effects before the data can be compared with the theoretical calculations.
However, adding the above systematic effects in quadrature, we get a 6\%
uncertainty on the measured distributions.  The small detector effects  suggest
that the data distributions can be directly compared with the  parton--level
distributions of perturbative QCD calculations. In the following, the  measured
distributions with a 6\% estimated total systematic error  are directly
compared with the QCD tree--level  calculations at the parton level.  Finally,
we note that changing the jet separation requirement $\Delta{\cal R}$ from 1.4
to 1.0 does not change the degree of agreement between the data and QCD
calculations.

\section*{The Topologies of Three--Jet Events}

Figure~\ref{fig:3jx35} and Table~\ref{tab:3jx35} show the measured $x_3$ and
$x_5$ distributions  for the final selection of three--jet events. The three
jets are labeled  in order of decreasing energy in their \cms\ frame.  The
average values of $x_3$ and  $x_5$ are 0.88 and 0.39 respectively. The data are
compared with the  predicted distributions of the exact QCD tree--level
calculations and the expectations from the phase--space model.  The QCD
calculations reproduce the measured  distributions well for the entire range.
Unlike the predictions of the phase--space model, the data heavily populate the
high $x_3$ region and have significant contributions at low  $x_5$ values, a
characteristic of gluon radiation. The decrease in $x_3$ distributions  at high
$x_3$ values is due to the $\Delta{\cal R}$ requirement in the event selection.
The bottom plot shows the fractional difference between  the data and the QCD
predictions with dotted lines indicating the estimated 6\% systematic error on
the measurement.  The differences between the data and the predictions are
generally within the  systematic error band. The root mean square~(RMS) of the
fractional differences between the data and the QCD predictions are 3.4\% for
$x_3$ and 3.9\% for $x_5$.

\begin{figure}[hbt]
  \begin{center}
    \begin{tabular}{ll}
      \psfig{figure=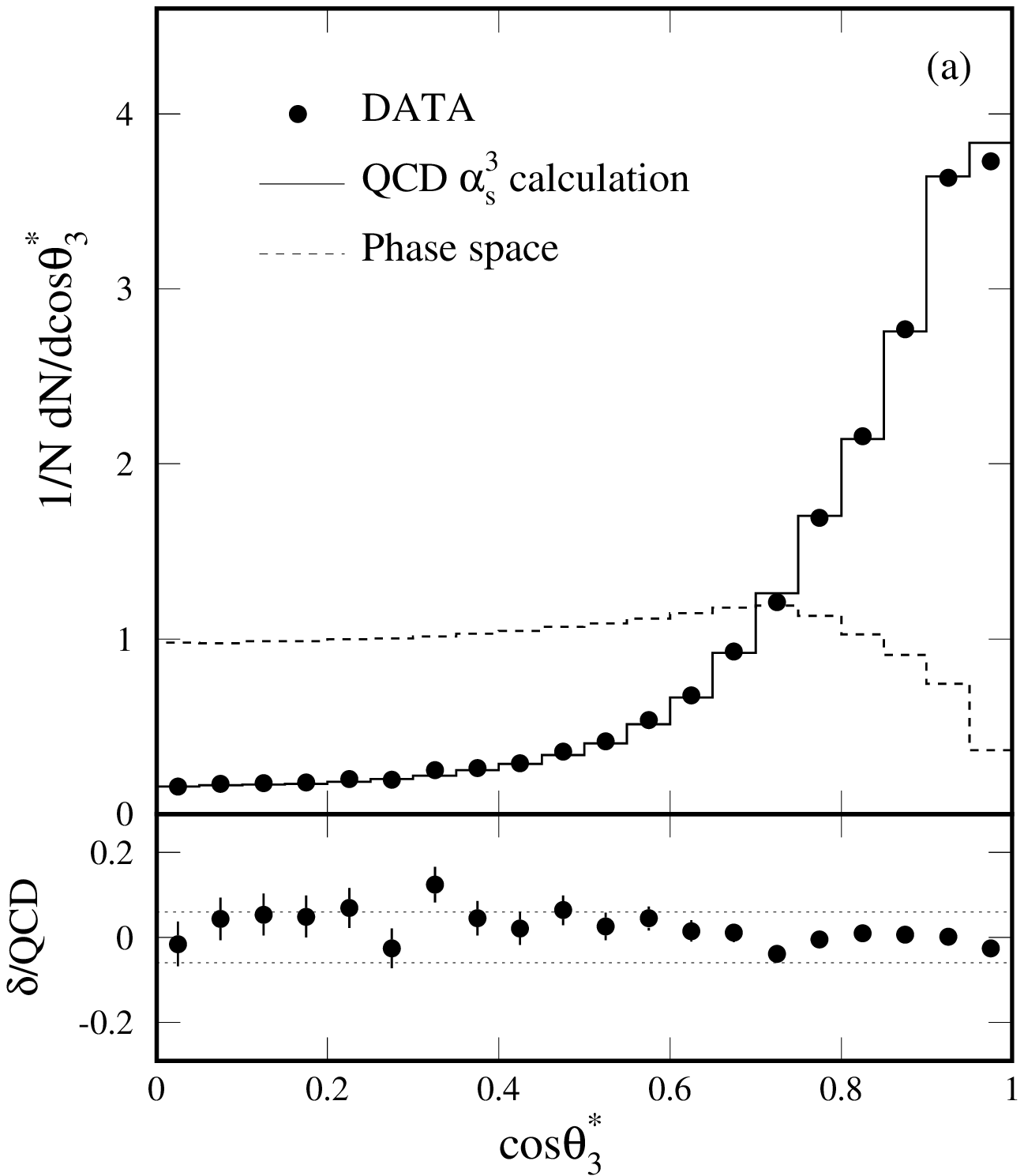,width=8.0cm} &
      \psfig{figure=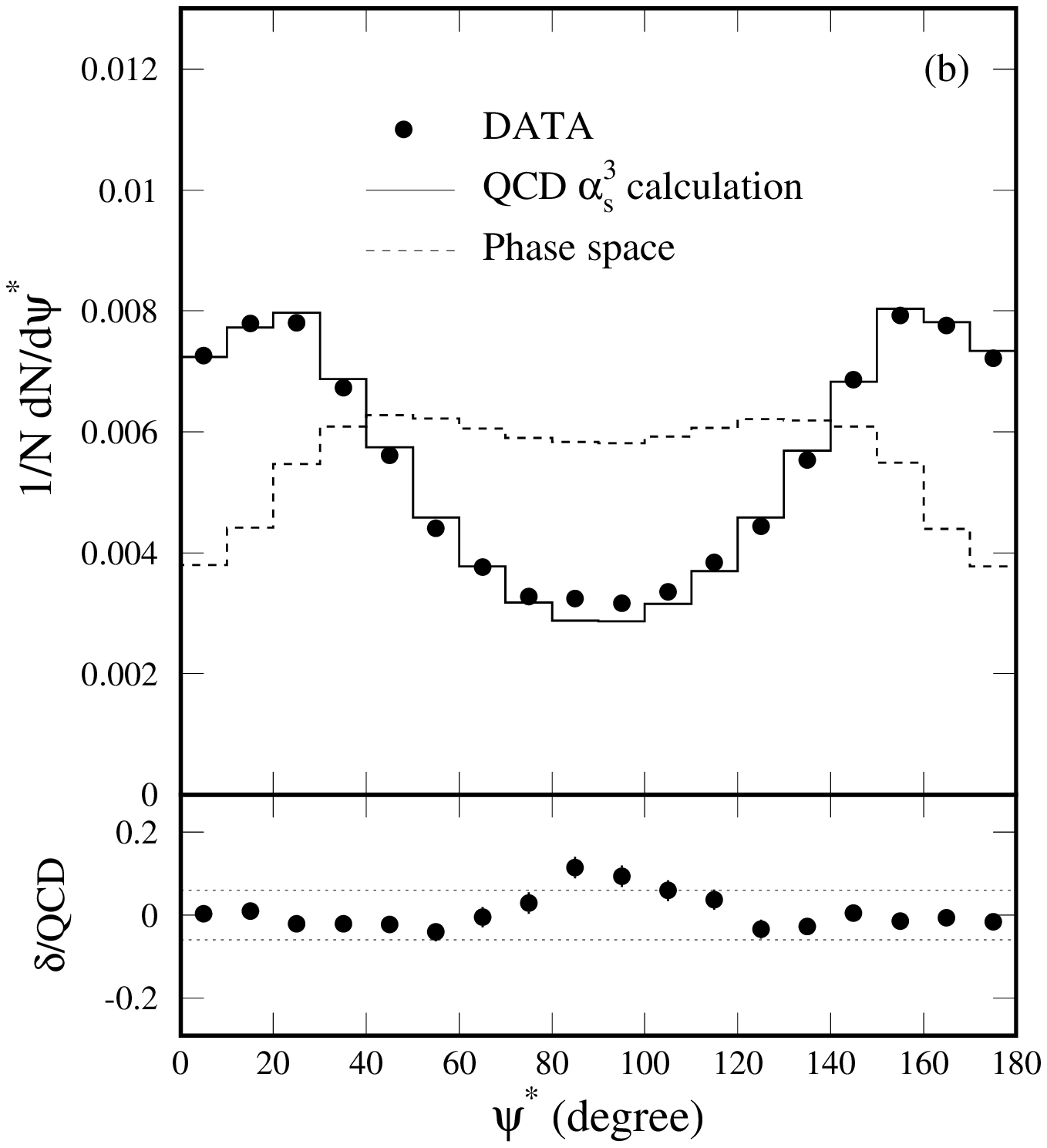,width=8.0cm} \\
    \end{tabular} \\
    \caption{The distributions for the three--jet events of (a) the cosine
              of the leading jet polar angle and (b) the angle $\psi^*$
              (defined in the text) in their center--of--mass system.
              The RMS of the fractional differences is 3.4\% for
              $\cos\theta^*_3$ and 4.2\% for $\psi^*$.
              The dotted lines show the
              estimated 6\% systematic uncertainty on the measurement.
              The data are also listed in Table~\protect\ref{tab:3jangle}.}
    \label{fig:3jangle}

\vspace*{1.0cm}
    \begin{tabular}{lcl}
      \psfig{figure=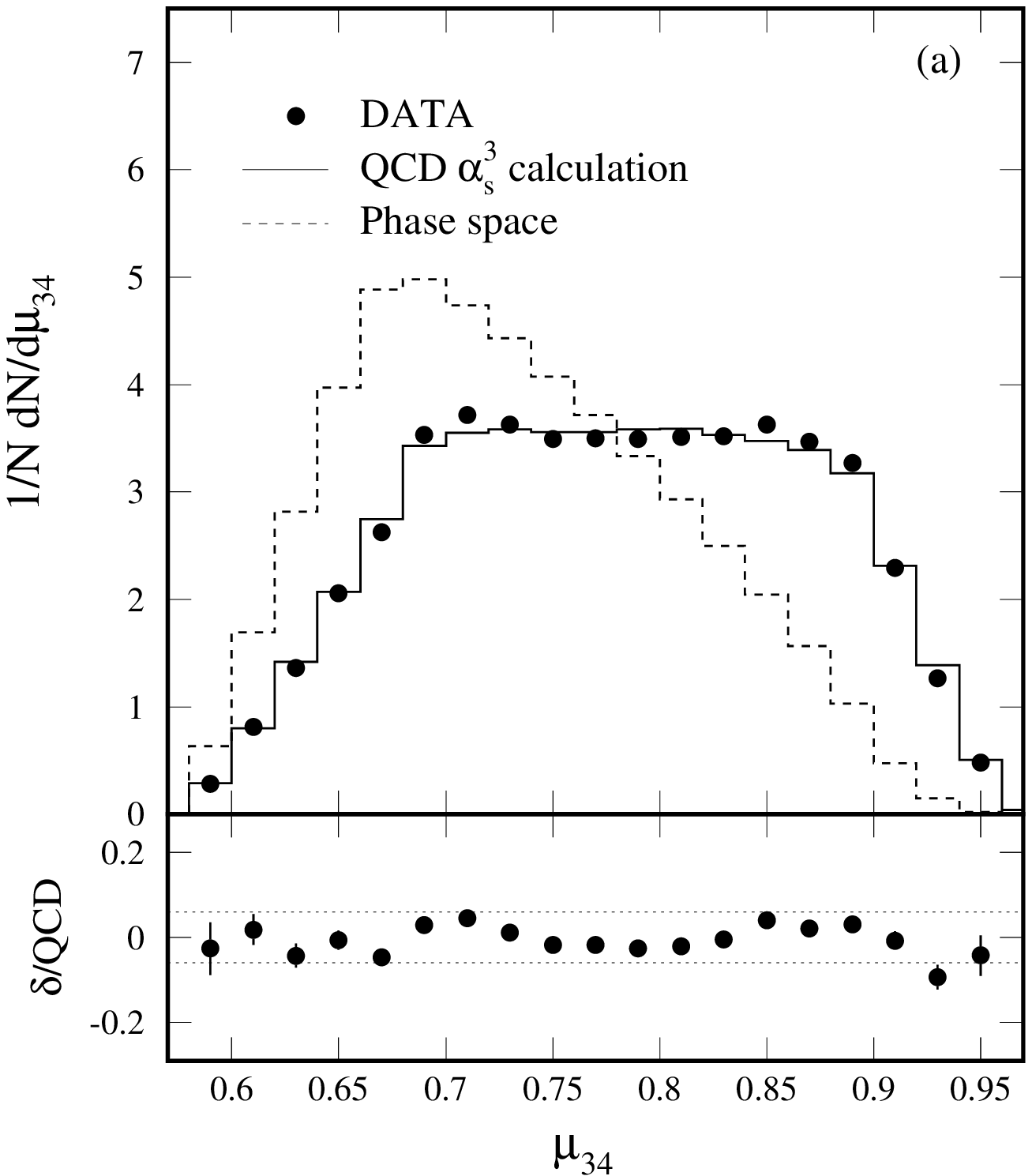,width=6.0cm} &
      \psfig{figure=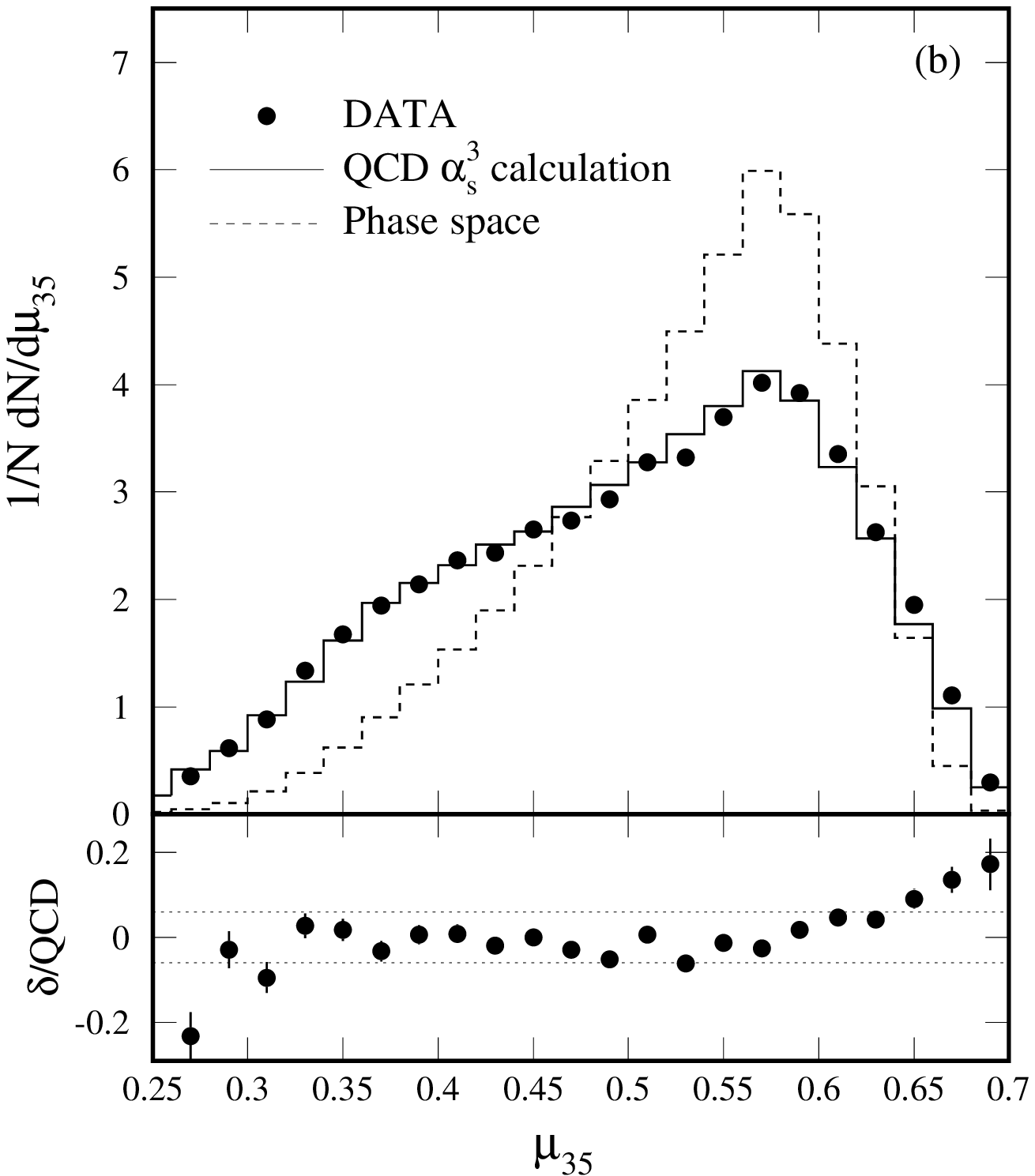,width=6.0cm} &
      \psfig{figure=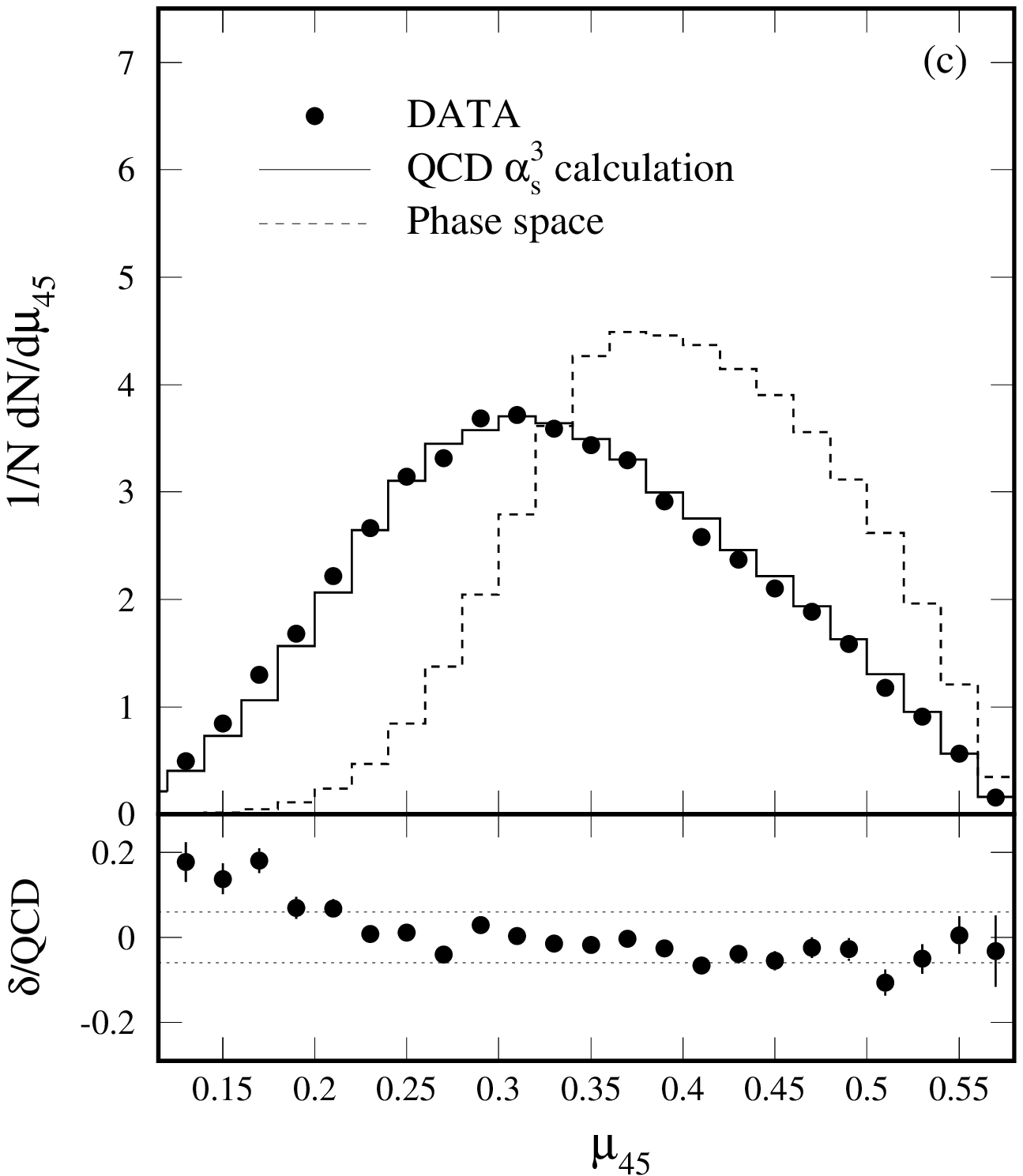,width=6.0cm} \\
    \end{tabular} \\
    \caption{The scaled mass distributions of (a) $\mu_{34}$, (b) $\mu_{35}$
             and (c) $\mu_{45}$ for the three--jet events in their
             center--of--mass system. The bottom plots show fractional
             differences between the data and QCD. The RMS's of the
             fractional differences are 3.6\%, 6.7\% and 6.9\%
             respectively. The dotted lines
             show the 6\% systematic uncertainty on the measurement.
             The data are also listed in Table~\protect\ref{tab:3jmass}.}
    \label{fig:3jmass}
  \end{center}
\end{figure}

The $\cos\theta^*_3$ distribution is shown in Fig.~\ref{fig:3jangle}(a).  As in
the angular distribution of two--jet events, an angular dependence
characteristic of Rutherford $t$--channel scattering is unmistakable.  The
large angular coverage of the D\O\ calorimeter allows the analysis to  cover
the entire $\cos\theta^*_3$ range, extending the study into  a previously
untested region of phase space. As is evident in the figure, the data are  well
reproduced by the predictions of the exact QCD tree--level calculations over
the entire range of $\cos\theta^*_3$, with a RMS 4.0\% of the fractional
differences. The phase--space distribution is mostly flat with high
$\cos\theta^*_3$ bins suppressed  as a result of the pseudorapidity requirement
in the  event selection. The depletion in the data and the QCD calculations is
compensated by a large cross section in this region and therefore is less
visible.  The measured $\psi^*$ distribution is shown in
Fig.~\ref{fig:3jangle}(b) together with the results of the exact QCD
tree--level calculation and of the phase--space model.  The phase--space
distribution shows depletions  at small and large $\psi^*$ angles, an effect of
the event selection. However, the data and the QCD distributions are enhanced
in these regions because of initial--state radiation in which one of the two
non--leading jets is close to the beam line.  As in the case of the $x_3$,
$x_5$ and $\cos\theta^*_3$ distributions, the overall agreement between data
and the QCD tree--level calculations is  very good. The RMS of the fractional
differences is 4.2\%.

The scaled mass distributions are sensitive to the jet energies, the opening
angles between jets, and the correlations between these quantities.  The
measured $\mu_{34},\ \mu_{35}$ and $\mu_{45}$ distributions for the three--jet
event sample are compared with  the exact QCD predictions in
Fig.~\ref{fig:3jmass}. The QCD predictions agree with the data well, while the
differences between the data and the phase--space model are large. We also note
that some systematic shift in $\mu_{35}$ and $\mu_{45}$ distributions is
clearly visible. The RMS's of the fractional differences between the data and
the QCD calculations are 3.6\%, 6.7\% and 6.9\% for  $\mu_{34}$, $\mu_{35}$ and
$\mu_{45}$ respectively.

Finally, we note that the KS approximate QCD calculations are essentially
identical to the exact tree--level QCD calculations for the topological
variables studied above. This implies that the topological distributions are
insensitive to the approximation made in the KS calculations.

\section*{The Topologies of Four--Jet Events}

\begin{figure}[hbt]
  \begin{center}
    \mbox{\psfig{figure=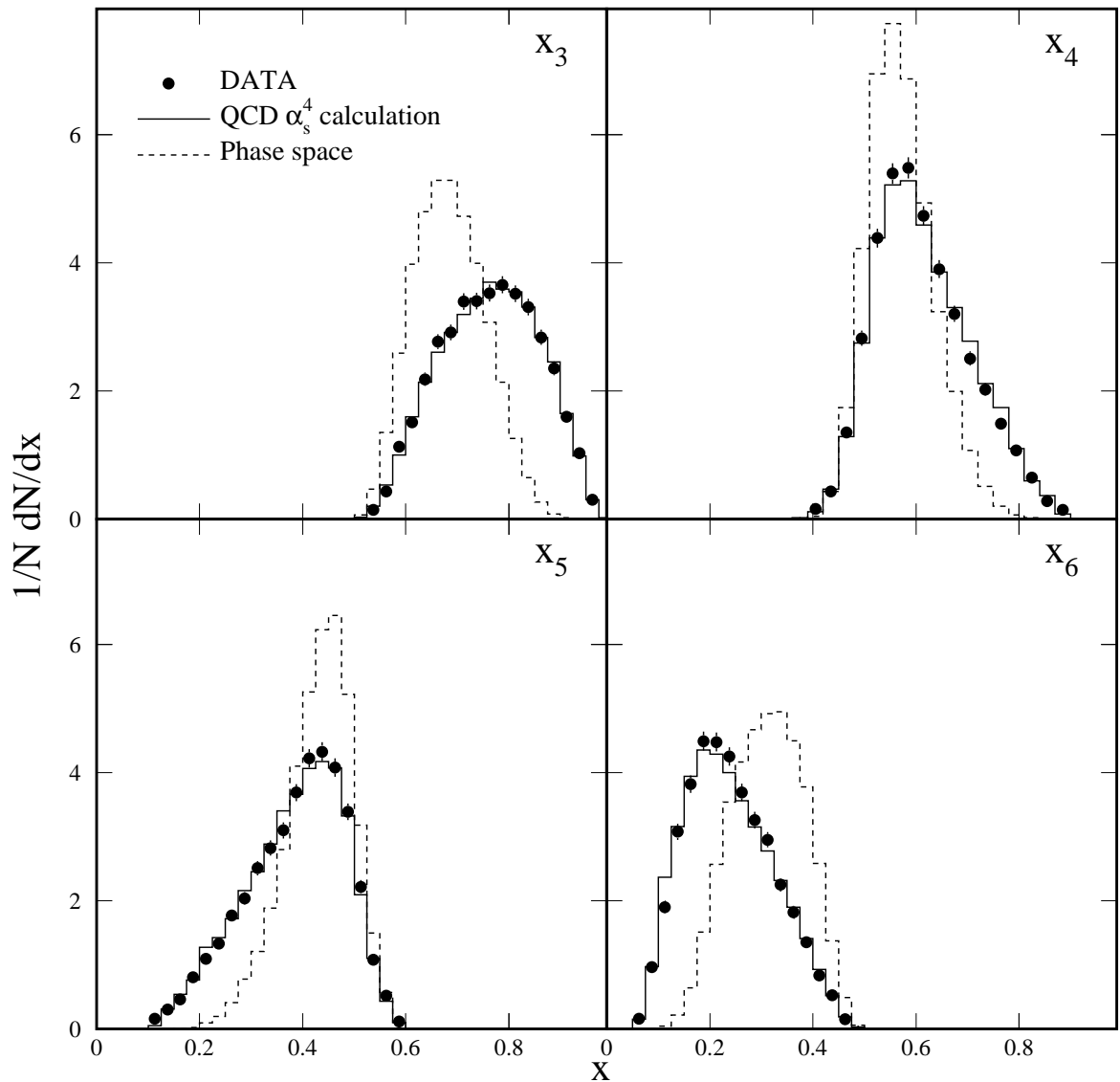,width=16.0cm}}
    \caption{The jet energy fraction distributions for the four--jet events
             in their center--of--mass system. Only statistical errors are
             shown. The RMS's of the fractional differences between the data
             and QCD are 6.5\%, 5.5\%, 6.2\% and 6.5\% for $x_3$, $x_4$,
             $x_5$ and $x_6$ respectively.
             The estimated systematic uncertainty on the measurement is 6\%.
             The data are also listed in Table~\protect\ref{tab:4jxvar}.}
    \label{fig:4jxvar}
  \end{center}
\end{figure}

The four measured energy fractions of four--jet events are shown in
Fig.~\ref{fig:4jxvar} and also listed in Table~\ref{tab:4jxvar}.  The four jets
are labeled in order of  decreasing energy in their center--of--mass system.
Although four scaled energy variables are shown, only three of these are
independent. The other is fixed by the condition $\sum_{i} x_i = 2$.  The
measured mean values of the four energy fractions are 0.76, 0.61, 0.39, and
0.24. The QCD predictions of the exact tree--level calculations are represented
by the solid curves and are in an excellent agreement  with the data for all
four variables. As in the three--jet case,  the distributions from the
phase--space model do not reproduce the data. The fractional differences
between the data and QCD are very similar to those of the three--jet events and
are not shown for simplicity.

\begin{figure}[hbt]
  \begin{center}
    \mbox{\psfig{figure=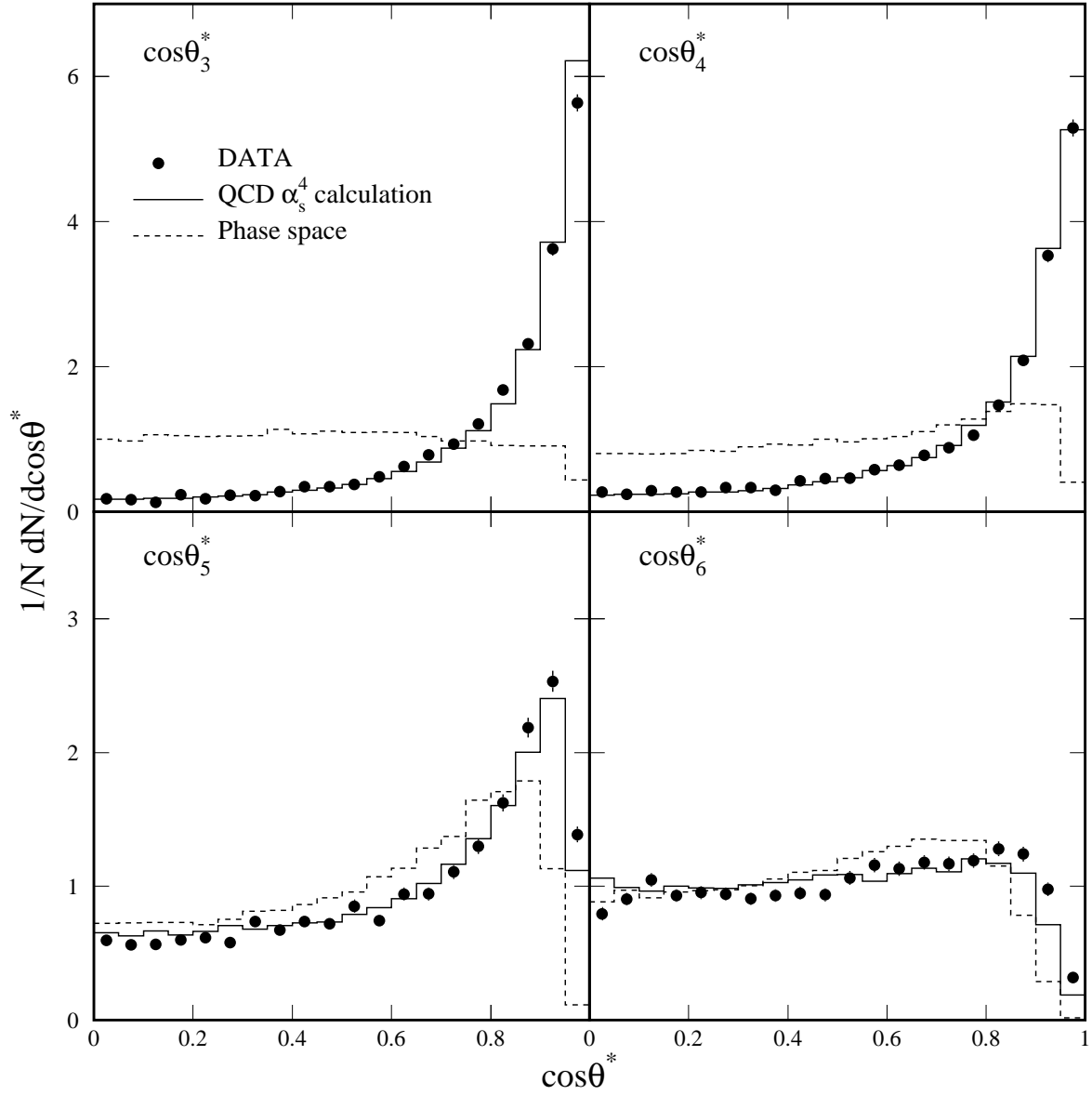,width=16.0cm}}
    \caption{The distributions of jet polar angle for the four--jet events
             in their center--of--mass system. Only statistical errors are
             shown. The RMS's of the fractional differences between the data
             and QCD are 6.8\%, 6.2\%, 7.5\% and 7.9\% for $\cos\theta^*_3$,
             $\cos\theta^*_4$, $\cos\theta^*_5$ and $\cos\theta^*_6$
             respectively.
             The estimated systematic uncertainty on the measurement is 6\%.
             The data are also listed in Table~\protect\ref{tab:4jcos}.}
    \label{fig:4jcos}
  \end{center}
\end{figure}

The cosines of the four polar angles of the four--jet events in their
center--of--mass system are compared with QCD calculations in
Fig.~\ref{fig:4jcos} for the entire range. While the two leading jets tend to
be in the forward region, the cosine distribution of the least energetic jet is
essentially flat, because  the jet separation requirement in the event
selection favors events with  other jets in the central region. Although small
differences  between the data and the QCD calculations are visible, the overall
agreement is good. Despite the large differences between the data and the
phase--space model in $\cos\theta^*_3$ and $\cos\theta^*_4$  distributions, the
differences in the other two distributions are relatively small.

\begin{figure}[hbt]
  \begin{center}
    \mbox{\psfig{figure=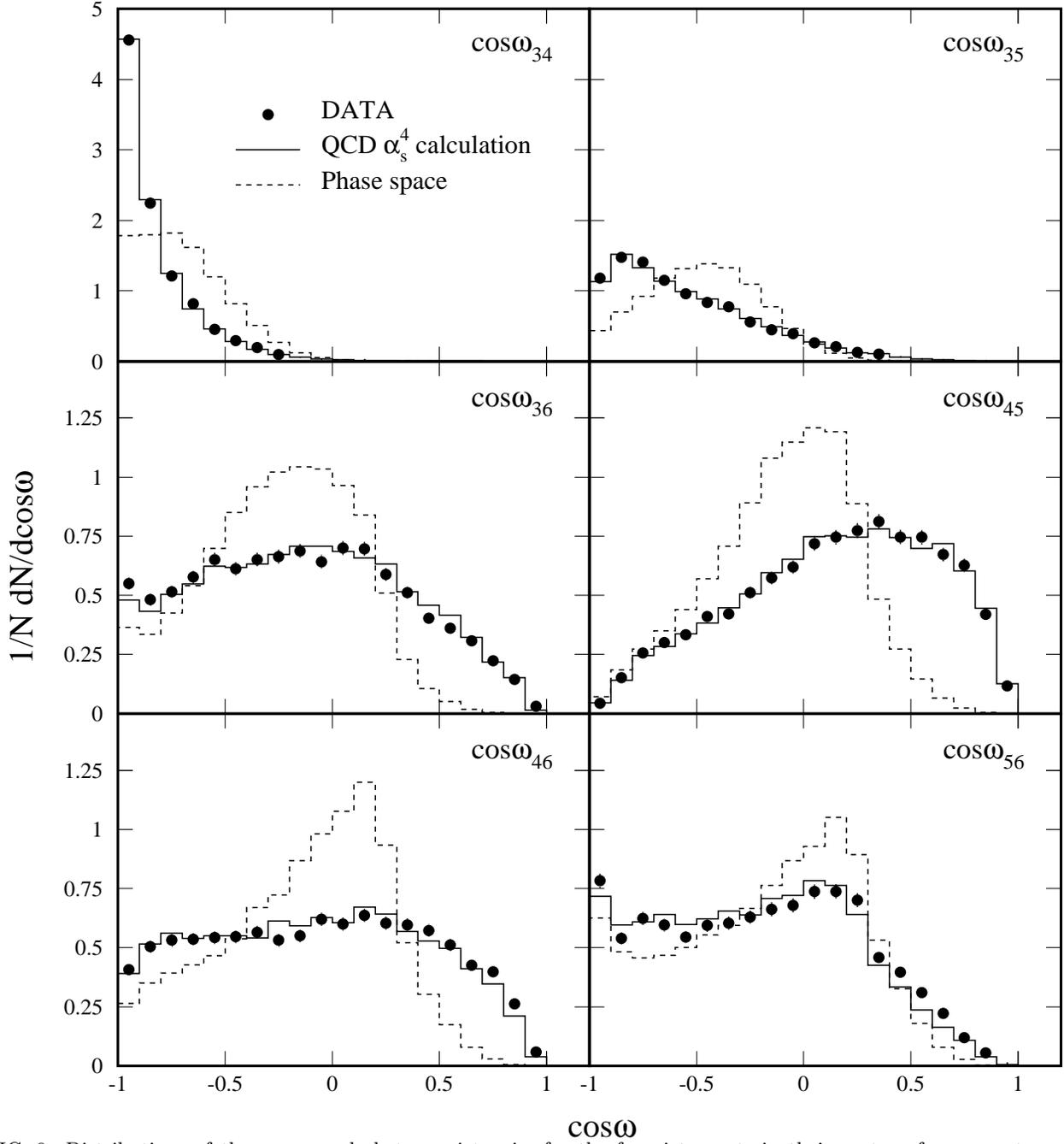,width=17.0cm}}
    \caption{Distributions of the space angle between jet pairs for the
             four--jet events in their center--of--mass system. Only
             statistical errors are shown. The RMS's of the fractional
             differences between the data and QCD for the
             six space angles are 5.4\%, 5.6\%, 6.4\%, 4.7\%, 7.6\% and 8.0\%.
             The estimated systematic uncertainty on the measurement is 6\%.
             The data are also listed in Table~\protect\ref{tab:4jangle}.}
    \label{fig:4jangle}
  \end{center}
\end{figure}

The internal event structure can be further understood by examining the opening
angles between jets. Figure~\ref{fig:4jangle} shows the   distributions of the
space angle between all  possible jet pairs of the four--jet events in their
center--of--mass system. While the two leading jets are mostly back--to--back,
the angles between other jet pairs are distributed widely. The depletion in the
regions where $\cos\omega_{ij}\rightarrow 1.0$ is again due to the $\Delta{\cal
R}$ requirement in the event selection. The structures of the data
distributions are well described by the QCD predictions.

\begin{figure}[hbt]
  \begin{center}
    \mbox{\psfig{figure=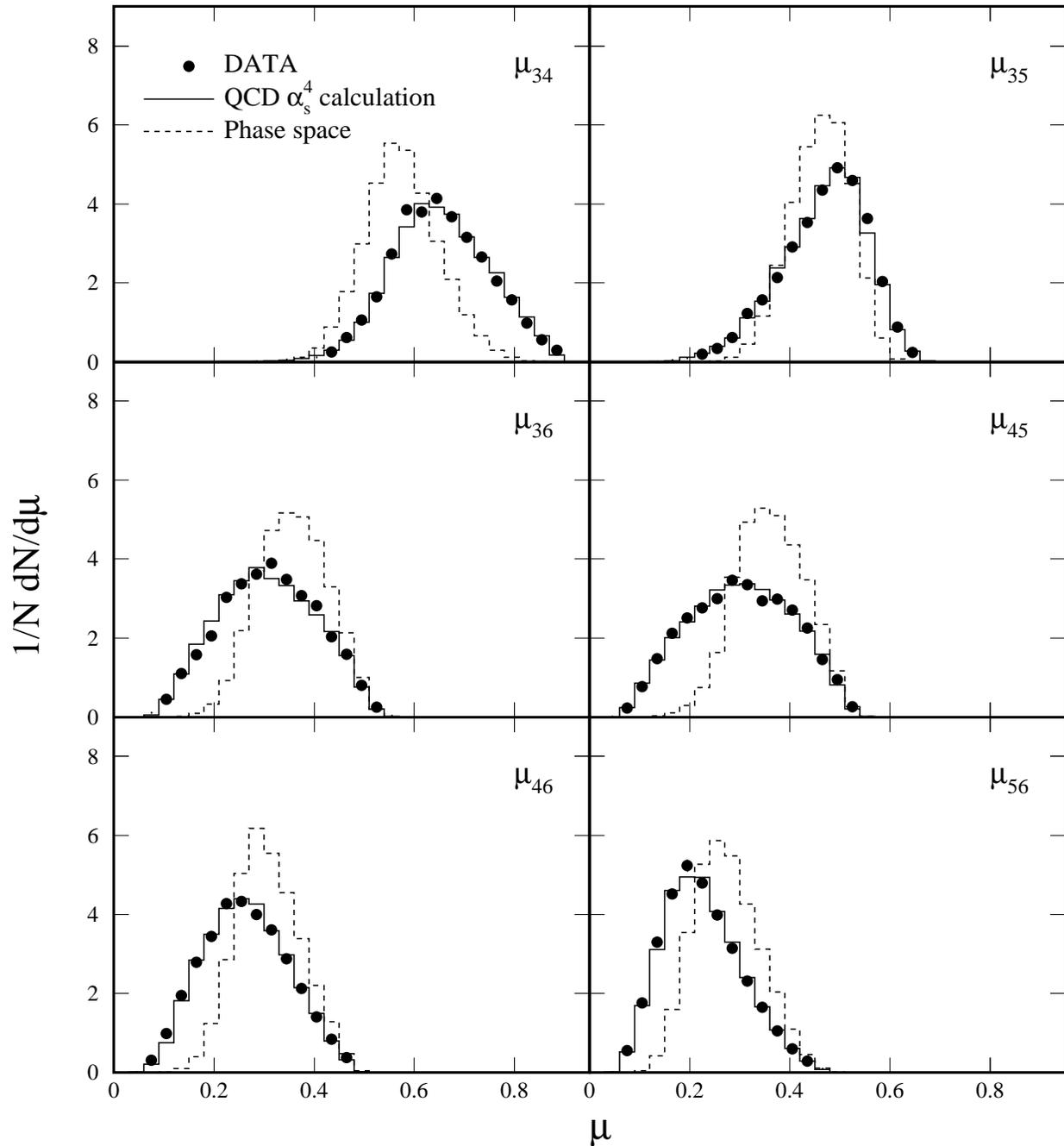,width=17.0cm}}
    \caption{Distributions of scaled jet pair mass for the four--jet events
             in their center--of--mass system. Only statistical errors are
             shown. The RMS's of the fractional differences between the data
             and QCD for
             the six masses are 7.8\%, 7.1\%, 6.8\%, 6.3\%, 4.1\% and 3.7\%.
             The estimated systematic uncertainty on the measurement is 6\%.
             The data are also listed in Table~\protect\ref{tab:4jmass}.}
    \label{fig:4jmass}
  \end{center}
\end{figure}

Figure~\ref{fig:4jmass} shows the scaled mass distributions of jet pairs of the
four--jet events for both data and the QCD calculations. The average scaled
mass is 0.65 for the two leading jets and is 0.23 for the two non--leading
jets. The QCD calculations agree with the data well. Distributions of the
phase--space model are generally too narrow and fail to reproduce the data
distributions.

\begin{figure}[hbt]
  \begin{center}
      \mbox{\psfig{figure=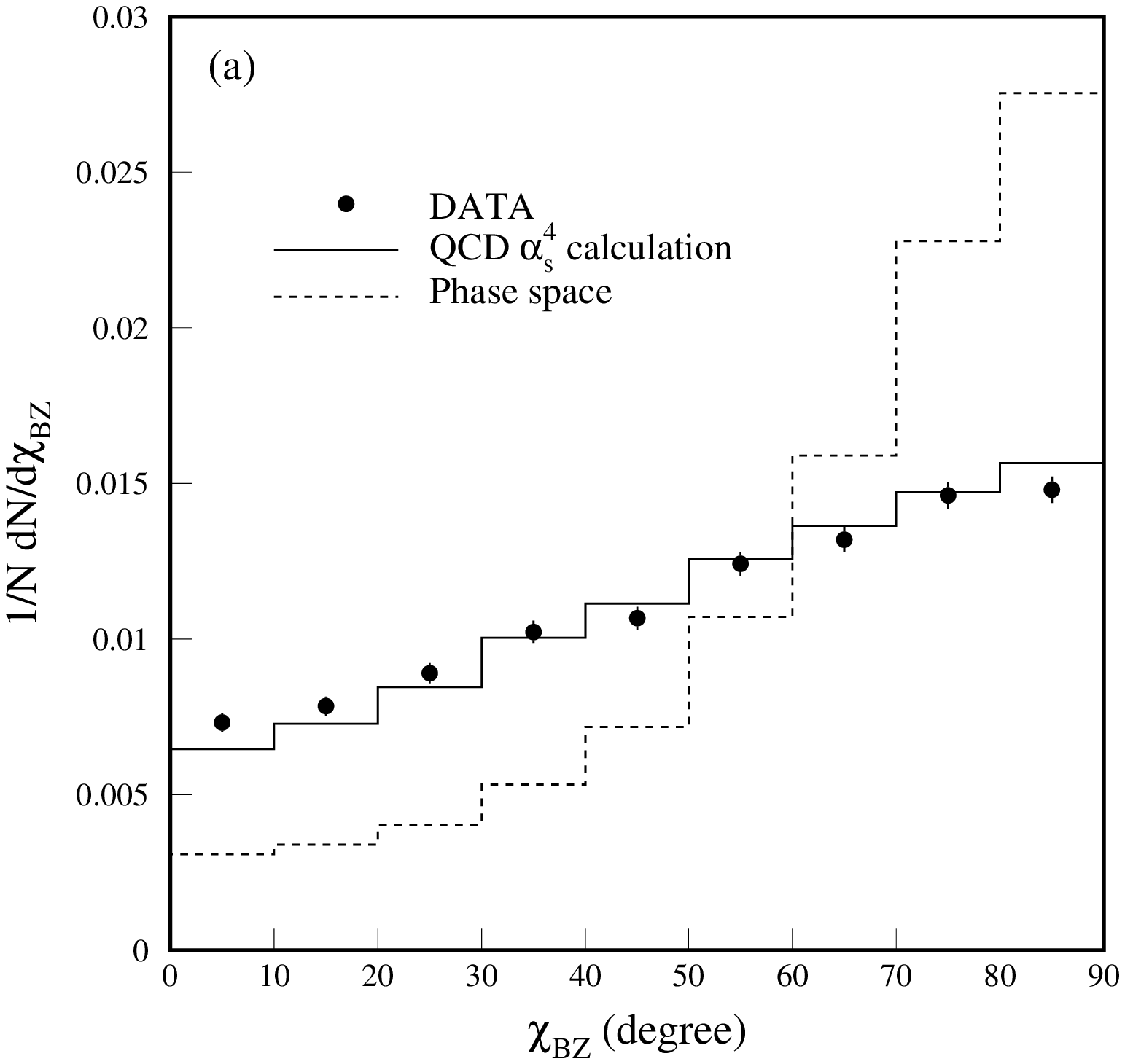,width=10.0cm}} \\
      \mbox{\psfig{figure=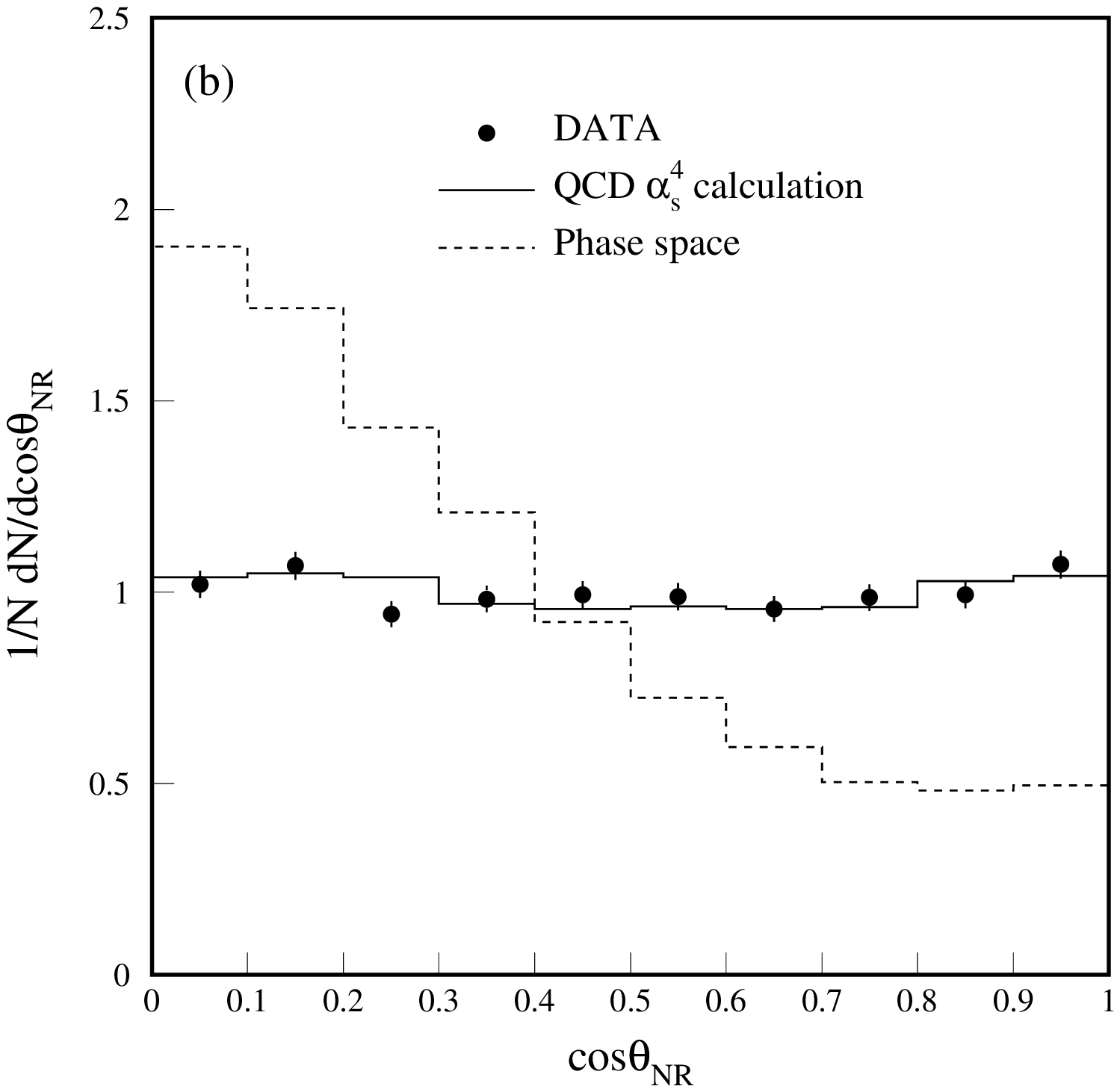,width=10.0cm}}
    \caption{ The distribution of the angle between the planes of (a) the
              two leading jets and the two non--leading jets and (b) the
              momentum vector differences of the two
              leading jets and the two non--leading jets, for four--jet
              events in their center--of--mass system.
              Only statistical errors are shown. The RMS of the fractional
              differences is 5.7\% for $\chi_{BZ}$ and 4.1\% for
              $\cos\theta_{NR}$. The estimated systematic uncertainty on
              the measurement is 6\%.
              The data are also listed in Table~\protect\ref{tab:4jbznr}.}
    \label{fig:4jbznr}
  \end{center}
\end{figure}

Figure~\ref{fig:4jbznr} compares the measured $\chi_{BZ}$ and $\cos\theta_{NR}$
distributions with the predictions of the exact tree--level QCD calculations as
well as those from the phase--space model. The agreement between the data and
QCD is generally good and the differences between the data and the phase--space
model are large. Although the jet separation requirement in the event selection
favors large $\chi_{BZ}$, the data and the QCD distributions have significant
contributions in the small $\chi_{BZ}$ region, which corresponds to a planar
topology of the four jets. In contrast, the phase--space distribution is highly
suppressed  in this region. The $\cos\theta_{NR}$ distributions for the data
and QCD are essentially flat while the phase--space model peaks strongly  as
$\cos\theta_{NR}$ approaches zero.

For the four--jet events as was the case for the three--jet events, the
normalized distributions from  the KS approximate QCD calculations agree well
with the data.

\section*{Comparison of QCD Subprocesses}

At the parton level, five and six partons (including the two initial partons)
are involved in the three-- and four--jet processes respectively. It is
difficult, if not impossible, to label quark or gluon jets in the data.
However, with Monte Carlo event generators, the three--jet cross section can be
broken into three subprocesses involving different numbers of quarks among the
initial-- or final--state partons: (1)~0--quark, (2)~2--quark and (3)~4--quark.
The predicted fractional contributions by \njets\ to the total three--jet
cross section for the selection criteria described above are 32.9\%, 50.8\% and
16.2\%  for 0--quark, 2--quark and 4--quark subprocesses respectively.
Similarly, the four--jet process can be divided into (1)~0--quark~(29.4\%),
(2)~2--quark~(49.6\%), (3)~4--quark~(20.2\%) and (4)~6--quark~(0.7\%)
subprocesses.

\begin{figure}[hbt]
  \begin{center}
    \begin{tabular}{ll}
      \psfig{figure=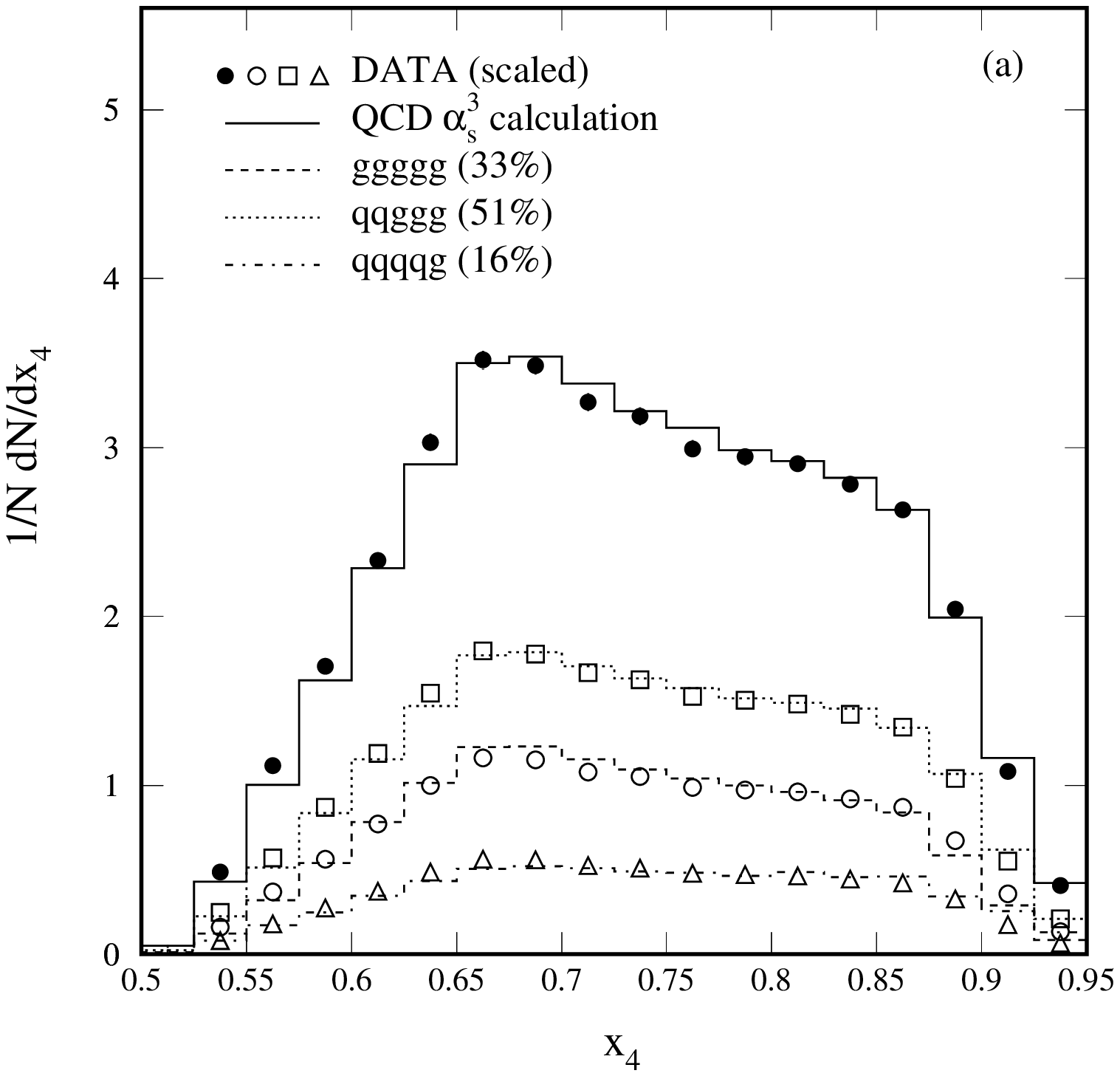,width=8.5cm} &
      \psfig{figure=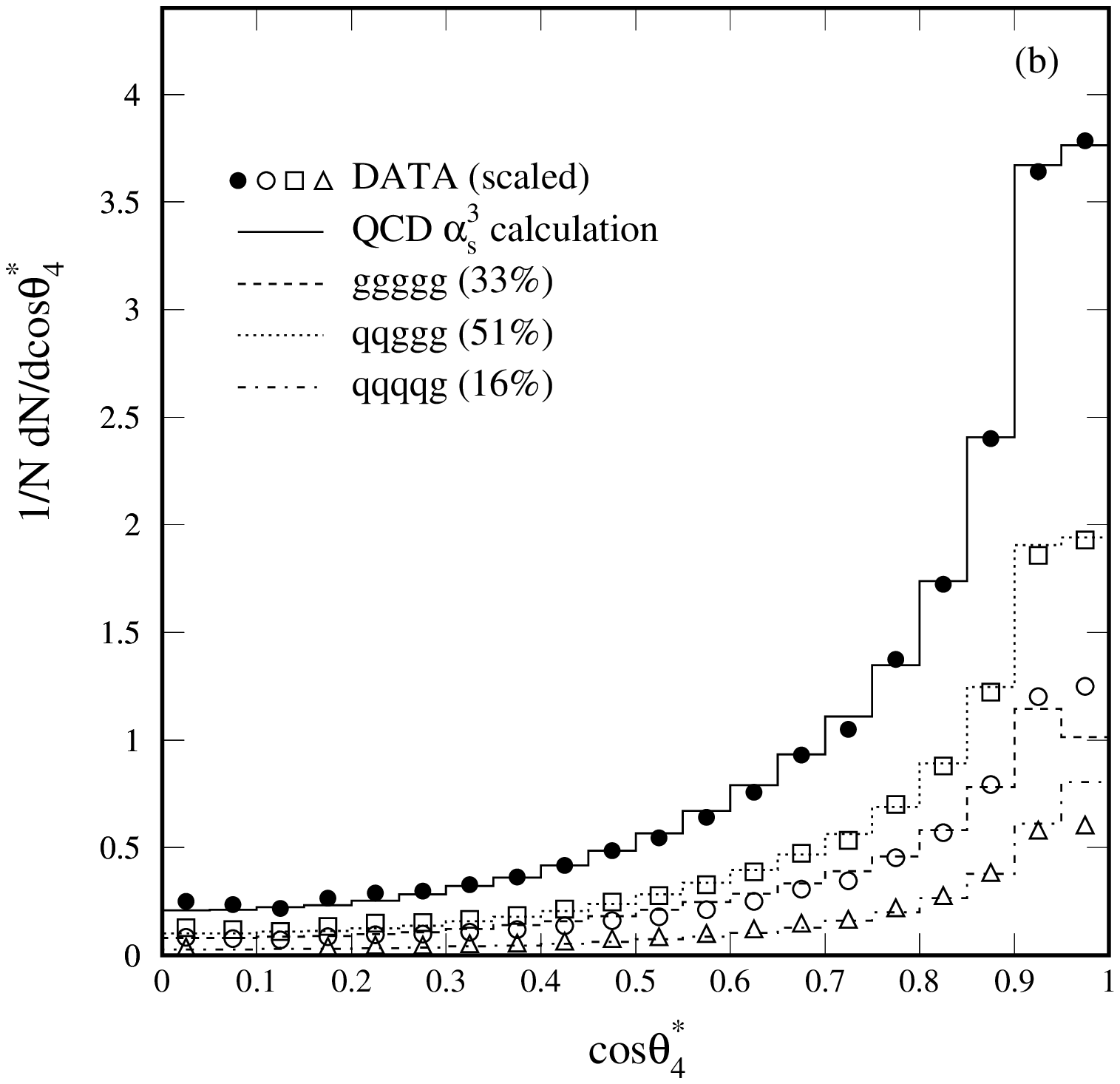,width=8.5cm} \\
      \psfig{figure=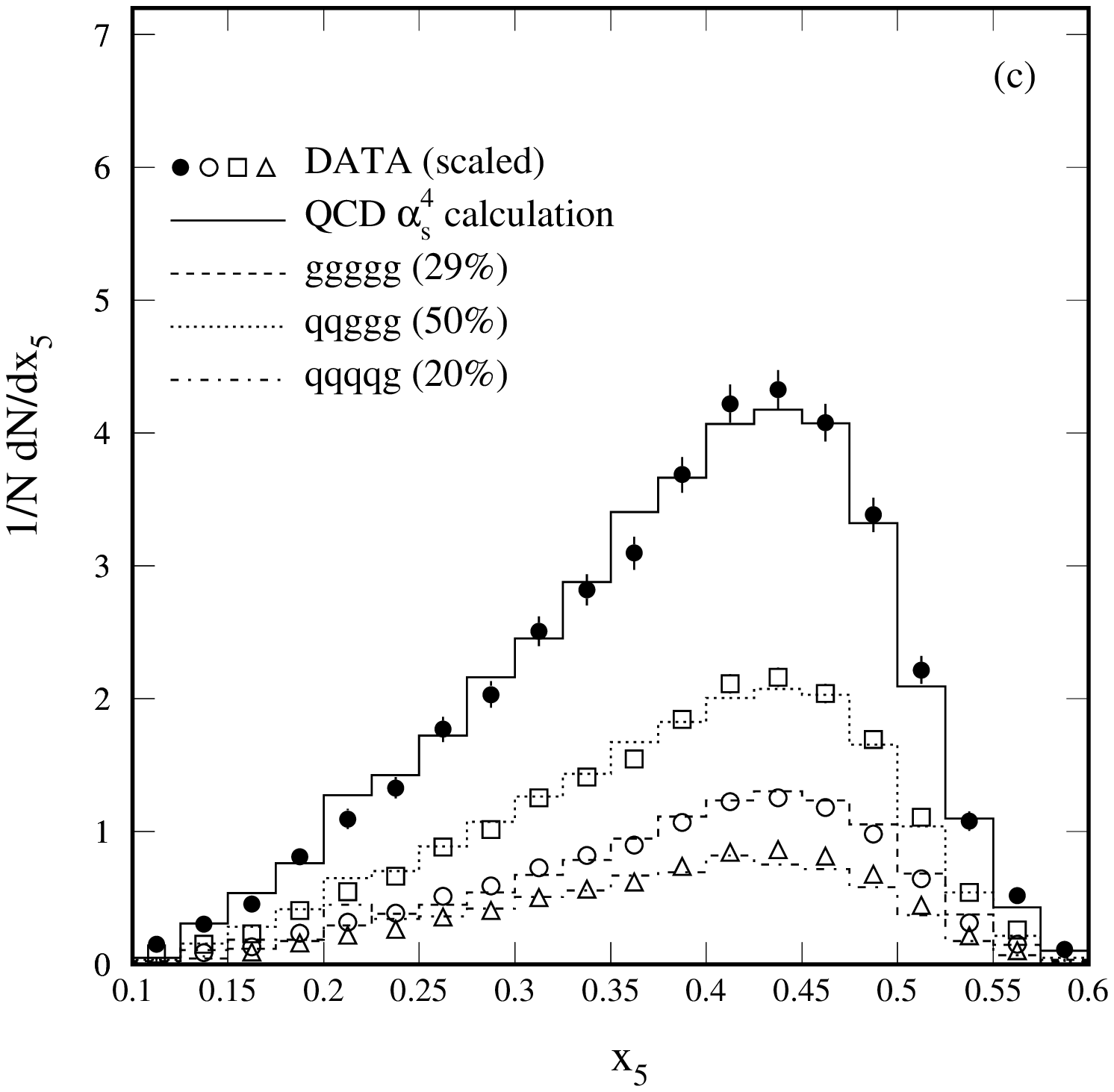,width=8.5cm} &
      \psfig{figure=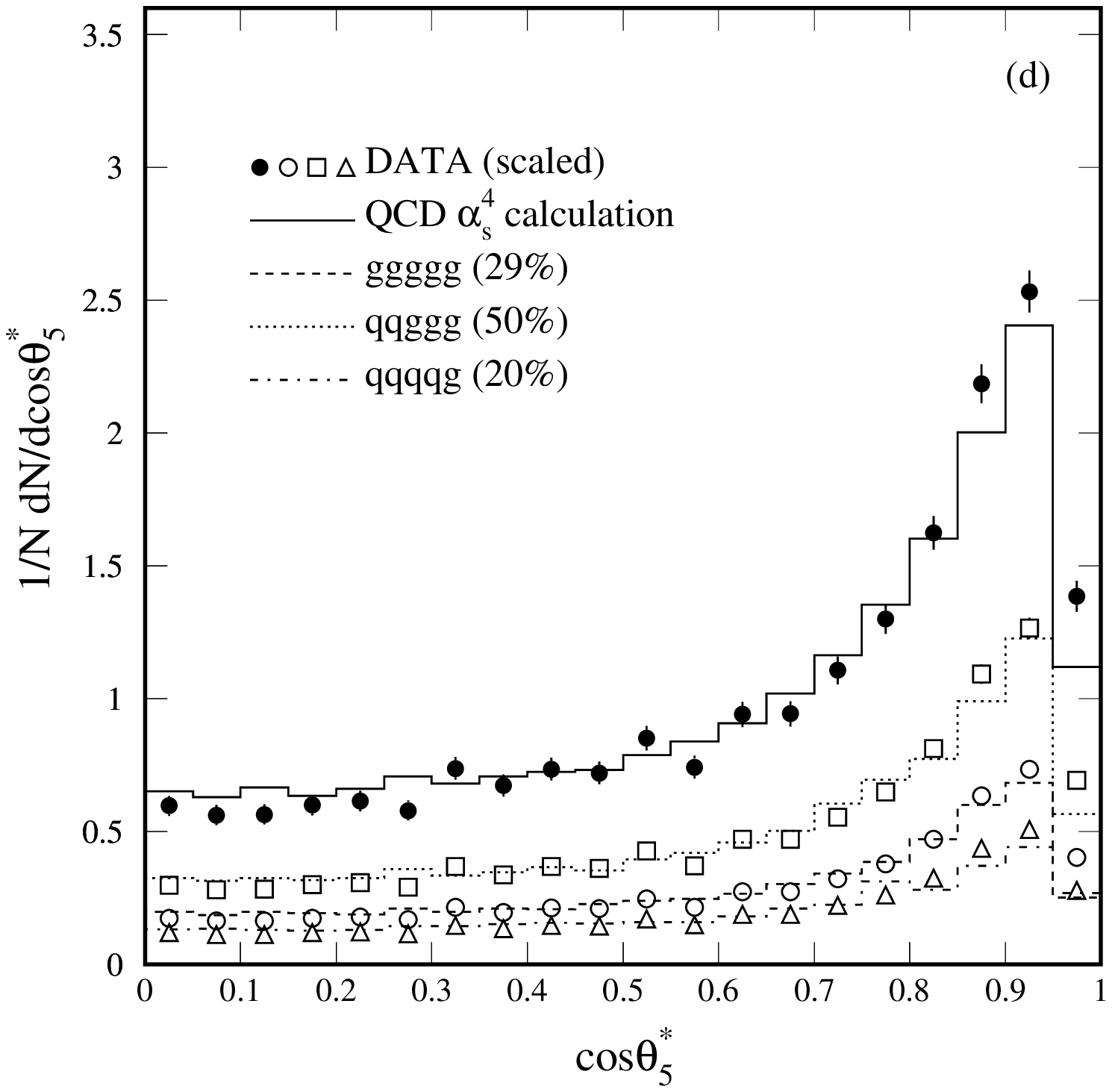,width=8.5cm} \\
    \end{tabular} \\
    \caption{The (a) $x_4$ and (b) $\cos\theta^*_4$ distributions for
             three--jet events and the (c) $x_5$ and (d) $\cos\theta^*_5$
             distributions
             for four--jet events in their center--of--mass system.
             The QCD subprocesses are normalized
             to their fractional contributions to the respective total
             cross section for the selection criteria described in the text.
             The data are scaled to the normalization of the respective
             subprocess of the QCD calculation. Therefore, only the
             shapes of the subprocesses are compared. The estimated uncertainty
             on the data is less than 6\%.}
    \label{fig:qg}
  \end{center}
\end{figure}

The studies described above show that the QCD calculations  agree well with the
data. It is therefore interesting to examine the topological distributions of
these subprocesses. Figures~\ref{fig:qg} (a) and (b) show the $x_4$ and
$\cos\theta^*_4$ distributions of the three--jet events  and Figs.~\ref{fig:qg}
(c) and (d) show the $x_5$ and $\cos\theta_5^*$ distributions of the four--jet
events predicted by the exact tree--level QCD calculations (full QCD) and by
the QCD calculations of the three subprocesses.  The full QCD is normalized to
unit area and the subprocesses are normalized to the fractional contribution to
their respective total cross section. The data distributions are normalized to
the respective QCD distributions. The distributions of the subprocesses are
remarkably similar and agree well with the data. The 6--quark subprocess
contributes less than 1\% of the total cross section of the four--jet events
and therefore is not shown in Figs.~\ref{fig:qg} (c) and (d). Nevertheless, the
normalized distributions are very similar to those of the other three
subprocesses. The similarity of the subprocesses is observed in all other
variables of the three-- and four--jet events investigated  in this paper.
This suggests that the distributions are insensitive to the relative
contributions of these subprocesses to the total cross section  and therefore
have weak dependences on the quark--gluon content  in parton distribution
functions.  Futhermore, Rutherford characteristics are visible in
$\cos\theta^*$  distributions for all subprocesses, implying that the matrix
elements of these subprocesses are dominated by the $t$--channel exchange.

\section*{Comparison with Parton Shower Event Generators}

As discussed above, the measured topological distributions of three-- and
four--jet events are reproduced well by the exact tree--level QCD calculations.
However, in many investigations, parton shower Monte Carlo event generators are
used to model multi--jet production. Therefore, it is interesting to compare
the data distributions with those predicted by parton shower event generators.

\begin{figure}[hbt]
  \begin{center}
    \begin{tabular}{ll}
      \psfig{figure=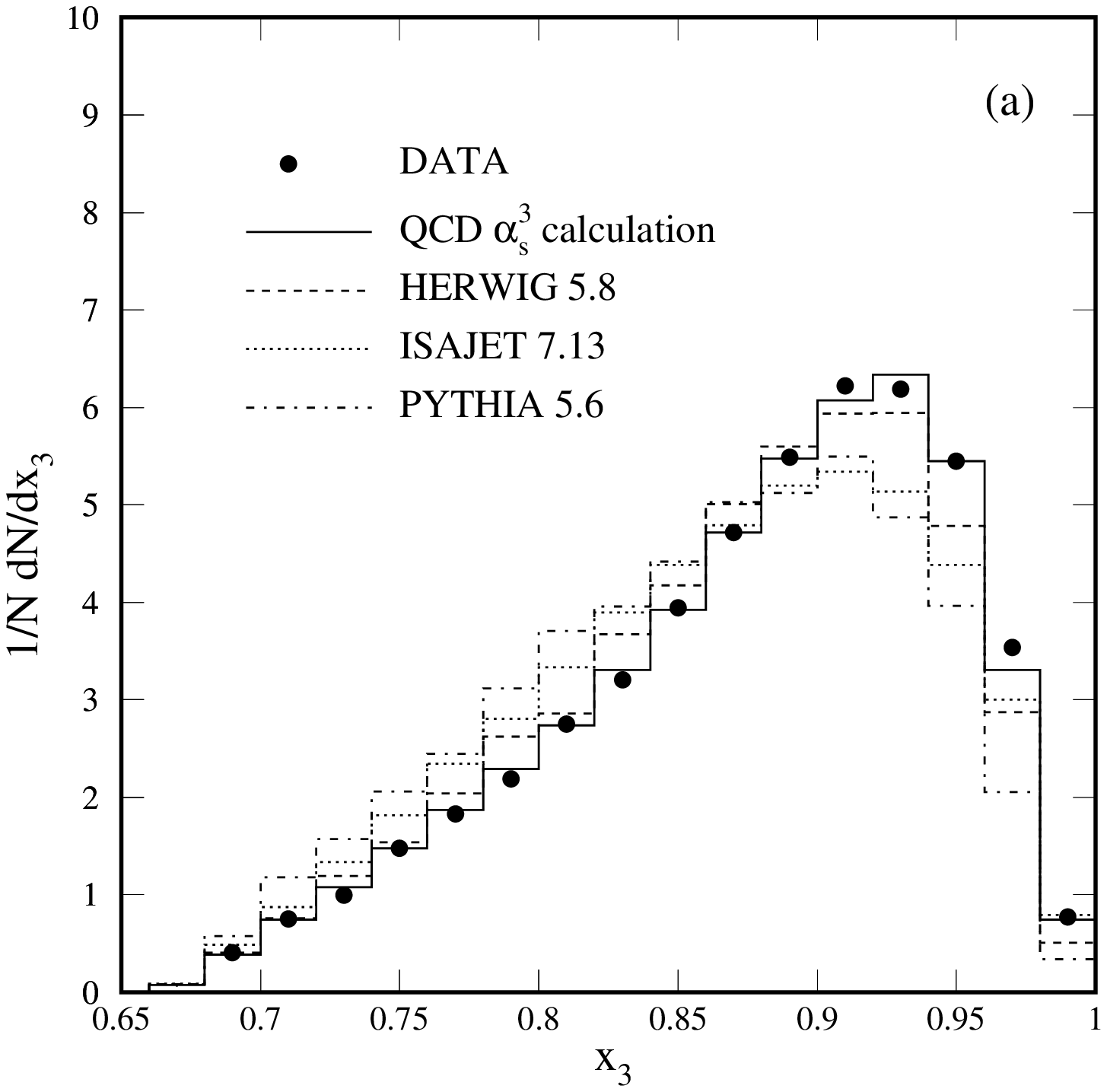,width=8.5cm} &
      \psfig{figure=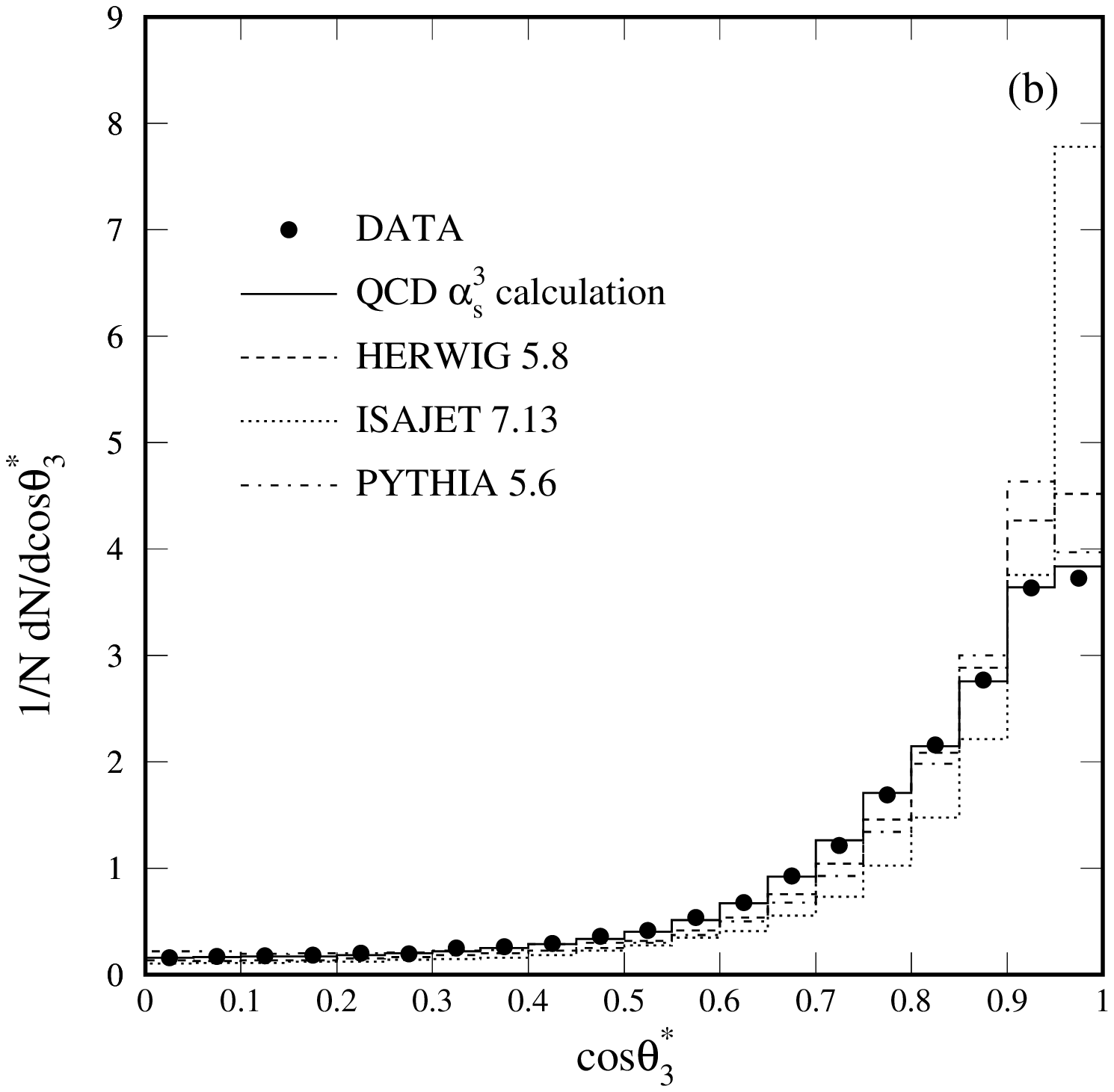,width=8.5cm} \\
      \psfig{figure=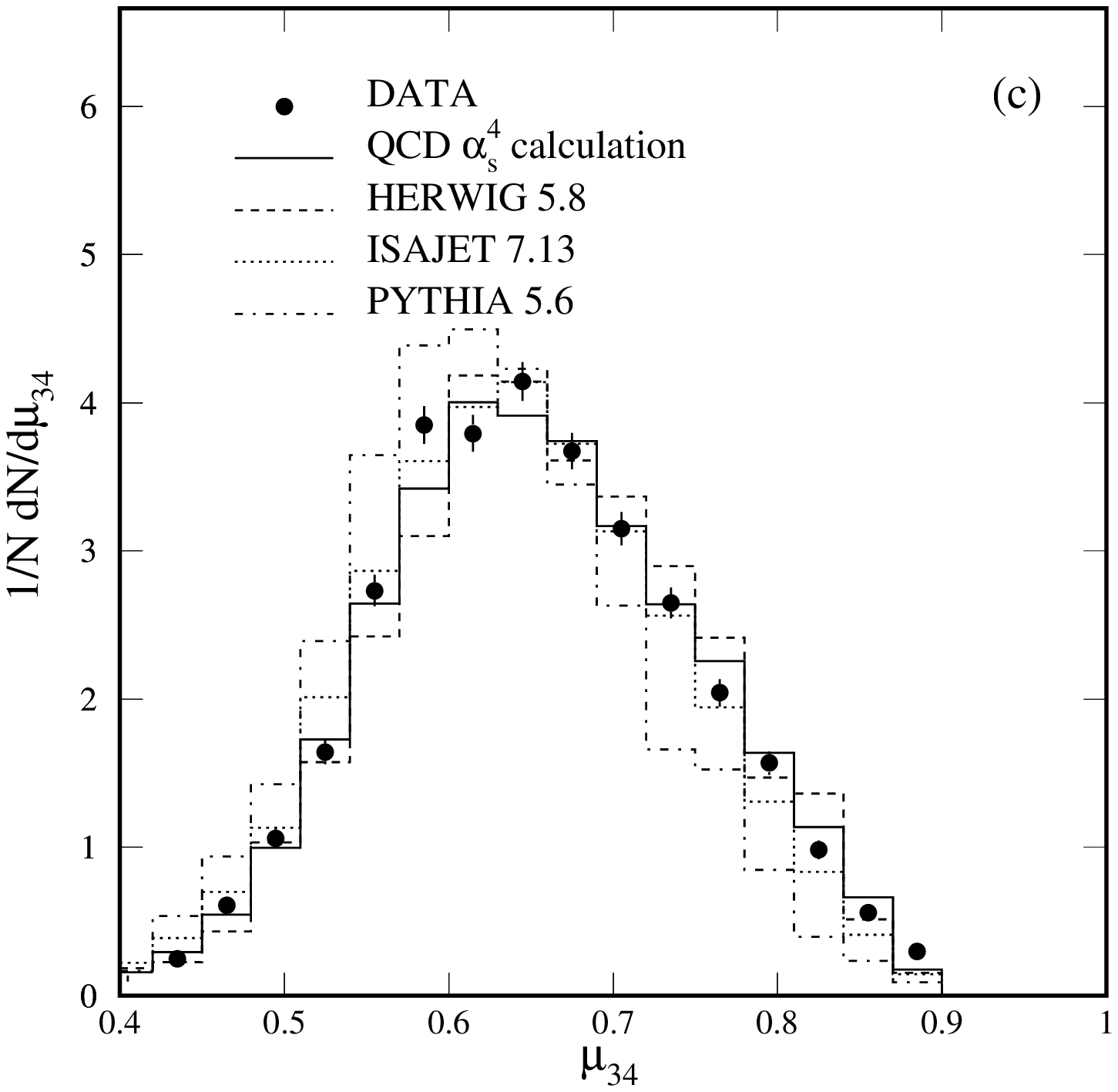,width=8.5cm} &
      \psfig{figure=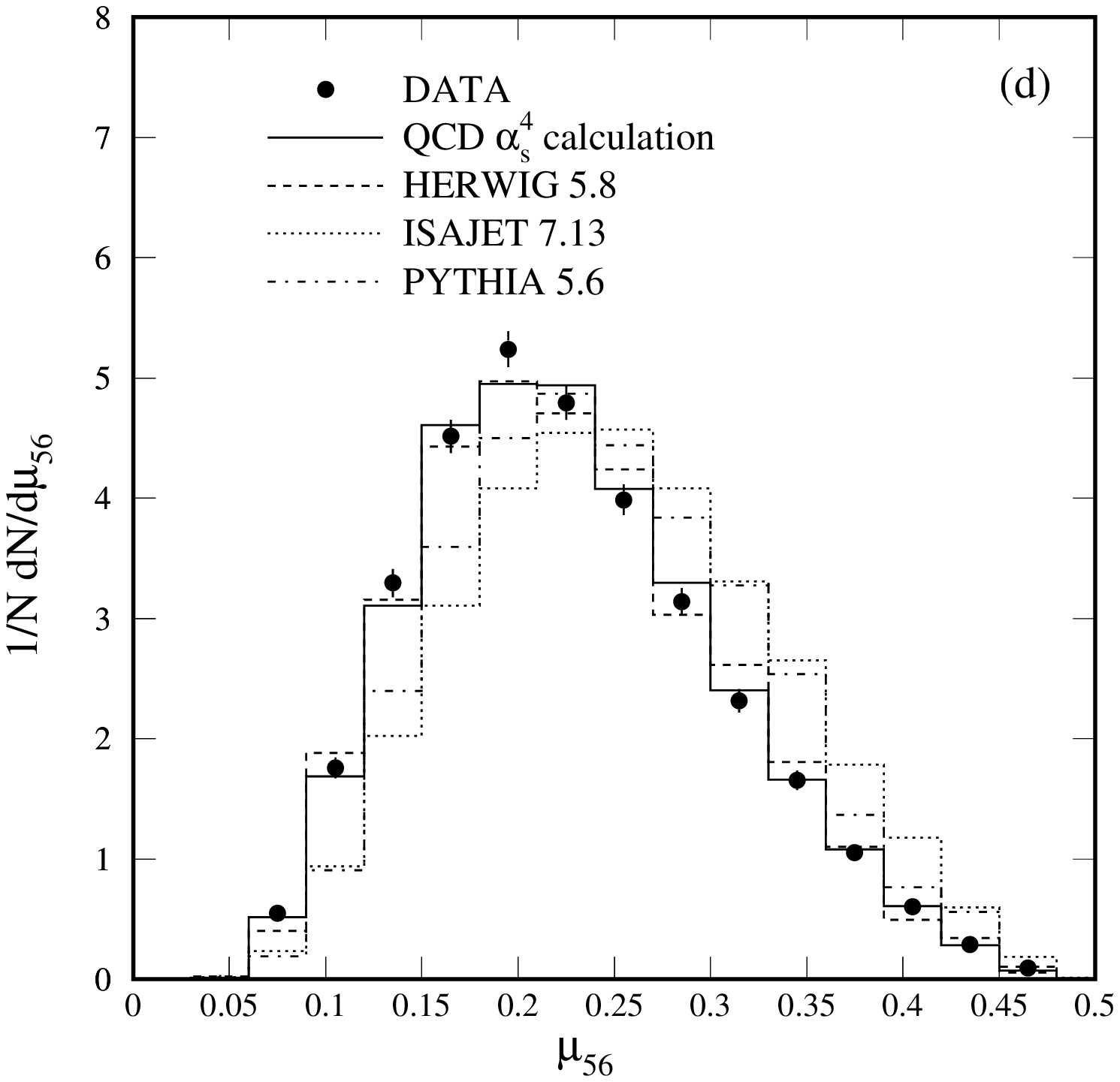,width=8.5cm} \\
    \end{tabular} \\
    \caption{Comparisons between the data, exact tree--level QCD calculations,
             and HERWIG, ISAJET and PYTHIA Monte Carlo predictions.  Shown
             are (a) the scaled
             energy of the leading jet and (b) the cosine of the leading jet
             for three--jet events, and the scaled invariant mass distributions
             of (c) the two leading jets and (d) the two non--leading jets
             for four--jet events. Only statistical errors are shown.
             The estimated systematic uncertainty on the
             measured distributions is less than 6\%.}
    \label{fig:ps}
  \end{center}
\end{figure}

As an example, the $x_3$ and $\cos\theta^*_3$ distributions of  three--jet
events and $\mu_{34}$ and $\mu_{56}$ distributions of four--jet events are
shown in Fig.~\ref{fig:ps} for the data and for the \HERWIG,  \ISAJET\ and
\PYTHIA\ parton shower event generators\footnote{All  parton shower events are
generated with  a $p_T=10$~GeV/c cutoff for the initial $2\rightarrow 2$ hard
scattering,  using their default parameters.}. The Monte Carlo distributions
are calculated using parton jets which are formed by quarks and gluons after
parton showering and before hadronization. The parton jets are initially
reconstructed using a cone jet algorithm implemented in the  \PYTHIA\
program~\cite{PYTHIA}. Then the jet direction is redefined  using a D\O\ jet
direction definition discussed above. Although the parton shower generators
describe the general structures of these variables well, differences in details
are  clearly visible. The largest difference is seen in the $\cos\theta_3^*$
distribution. All three parton shower event generators  show excessive
contributions in the forward region.

To generate three-- and four--jet events using the parton shower generators,
one has to start with $2\rightarrow 2$ processes with a $p_T$ cut and select
events with hard gluon radiation.  We note that a large fraction of the Monte
Carlo events in the forward region which pass the 60~GeV leading jet $E_T$
requirement have $2\rightarrow 2$ process with $p_T<50$~GeV/c. Presumably the
leading jets of the these events are from hard initial--state radiation.  This
observation suggests that the initial--state radiation is  not well modeled by
these parton shower generators in the phase--space region  studied in this
paper.

Although only four topological distributions are shown here, we have  compared
all other variables investigated in this paper.  Apart from the $\cos\theta^*$
distributions,  the \herwig\ event generator provides a reasonably good
description of the data  while the differences between the data and the
predictions of \isajet\ and \pythia\ event generators are large in many
distributions. Overall, the \herwig\ event  generator describes the data better
than the \isajet\ and the \pythia\ do.

\section*{Summary}

{}From the data sample recorded by the D\O\ detector  in $\pp$ collisions at
$\sqrt{s} = 1800$~GeV at the Tevatron during the 1992--1993 running period,
high statistics three--jet and four--jet event samples have been selected. A
large number of distributions characterizing the global structures of the
inclusive three-- and four--jet events have been compared with QCD calculations
of the exact tree--level matrix  elements and with calculations of QCD
subprocesses involving  different numbers of quarks. This paper extends earlier
studies to previously untested regions of phase space for a large number of
topological variables. For example, compared with an earlier study~\cite{CDF3}
of the three-jet events,  the $\cos\theta_3^*$ region studied has been expanded
from 0.8 to 1.0, the $x_3$ uplimit from 0.9 to 1.0 and the $\psi^*$ range from
$20^\circ<\psi^*<160^\circ$ to $0^\circ \le\psi^*\le 180^\circ$ for a minimum
three-jet invariant mass of 200~GeV/c$^2$. All comparisons have been made with
the parton--level distributions and based on normalized distributions rather
than cross sections.

For the three--jet events, the investigated topological variables are:  the
energy fractions carried by the two leading jets, the cosine of  the leading
jet polar angle, the angle between the plane containing  the leading jet and
the beam line, the plane containing the two non--leading jets, and the scaled
invariant masses of the jet pairs. In the case of the four--jet events, the
energy fractions and the cosines of the polar angles of all four jets, the six
opening angles, scaled invariant masses of jet pairs, and the angles between
jet planes have been studied.

Studies show that the measured topological distributions of the three--  and
four--jet events are well reproduced by the exact tree--level matrix  elements
QCD calculations. The good agreement implies that the topological
distributions of the three-- and four--jet events are determined by the
tree--level diagrams and therefore the topological distributions are not very
sensitive to higher--order corrections.  Futhermore, the distributions are
found  to be insensitive to the uncertainties in parton distribution functions
and  to the quark/gluon flavor of the underlying partons. The dominance of the
$t$--channel gluon exchange to a large extent determines the structure of the
event. The differences between the data and the phase space model are large for
most of the distributions. The successful direct comparison between  the data
and the QCD calculations at the parton level reaffirms the  assumption that
jets closely follow their underlying partons at high energies. Finally, we note
that apart from the $\cos\theta^*$ distributions, the \HERWIG\  event generator
provides a good description of the measured distributions while the differences
between the data and the predictions of the \ISAJET\ and the \PYTHIA\ event
generators are relatively large in many distributions.

\section*{ACKNOWLEDGMENTS}

We thank the Fermilab Accelerator, Computing, and Research Divisions, and the
support staffs at the collaborating institutions for their contributions to the
success of this work.   We also acknowledge the support of the U.S. Department
of Energy, the U.S. National Science Foundation, the Commissariat \`a L'Energie
Atomique in France, the Ministry for Atomic Energy and the Ministry of Science
and Technology Policy in Russia, CNPq in Brazil, the Departments of Atomic
Energy and Science and Education in India, Colciencias in Colombia, CONACyT in
Mexico, the Ministry of Education, Research Foundation and KOSEF in Korea and
the A.P. Sloan Foundation.

\newpage

\newpage
 \begin{table}[hbt]
   \begin{center}
      \begin{tabular}{|c|c||c|c|}\hline
   $x_3$  & $1/N dN/dx_3$ & $x_5$  & $1/N dN/dx_5$ \\ \hline
0.66$-$0.68 & 0.069$\pm$0.009          &  0.075$-$0.100 & 0.155$\pm$0.012
\\ 0.68$-$0.70 & 0.409$\pm$0.021 &  0.100$-$0.125 & 0.417$\pm$0.019
\\ 0.70$-$0.72 & 0.751$\pm$0.028 &  0.125$-$0.150 & 0.680$\pm$0.024
\\ 0.72$-$0.74 & 0.997$\pm$0.033 &  0.150$-$0.175 & 1.125$\pm$0.031
\\ 0.74$-$0.76 & 1.480$\pm$0.040 &  0.175$-$0.200 & 1.488$\pm$0.036
\\ 0.76$-$0.78 & 1.828$\pm$0.044 &  0.200$-$0.225 & 1.881$\pm$0.040
\\ 0.78$-$0.80 & 2.190$\pm$0.049 &  0.225$-$0.250 & 1.990$\pm$0.041
\\ 0.80$-$0.82 & 2.752$\pm$0.055 &  0.250$-$0.275 & 2.051$\pm$0.042
\\ 0.82$-$0.84 & 3.203$\pm$0.059 &  0.275$-$0.300 & 2.120$\pm$0.043
\\ 0.84$-$0.86 & 3.946$\pm$0.065 &  0.300$-$0.325 & 2.116$\pm$0.043
\\ 0.86$-$0.88 & 4.714$\pm$0.071 &  0.325$-$0.350 & 2.151$\pm$0.043
\\ 0.88$-$0.90 & 5.488$\pm$0.077 &  0.350$-$0.375 & 2.185$\pm$0.043
\\ 0.90$-$0.92 & 6.225$\pm$0.082 &  0.375$-$0.400 & 2.268$\pm$0.044
\\ 0.92$-$0.94 & 6.189$\pm$0.082 &  0.400$-$0.425 & 2.251$\pm$0.044
\\ 0.94$-$0.96 & 5.452$\pm$0.077 &  0.425$-$0.450 & 2.326$\pm$0.045
\\ 0.96$-$0.98 & 3.537$\pm$0.062 &  0.450$-$0.475 & 2.468$\pm$0.046
\\ 0.98$-$1.00 & 0.770$\pm$0.029 &  0.475$-$0.500 & 2.550$\pm$0.047
\\ & &  0.500$-$0.525 & 2.658$\pm$0.048       \\ &
&  0.525$-$0.550 & 2.314$\pm$0.045       \\ &                       &
0.550$-$0.575 & 1.848$\pm$0.040       \\ &                       &
0.575$-$0.600 & 1.378$\pm$0.035       \\ &                       &
0.600$-$0.625 & 0.957$\pm$0.029       \\ &                       &
0.625$-$0.650 & 0.508$\pm$0.021       \\ &                       &
0.650$-$0.675 & 0.090$\pm$0.009       \\ \hline\end{tabular}  \caption{The
measured $x_3$ and $x_5$ distributions with their statistical errors for the
three--jet events. The estimated systematic uncertainty is 6\%.}
\label{tab:3jx35}

\vspace*{1.5cm}
      \begin{tabular}{|c|c||c|c|} \hline
   $\cos\theta^*_3$  & $1/N dN/d\cos\theta^*_3$ & $\psi^*$
   & $1/N dN/d\psi^*$ \\ \hline
0.00$-$0.05 & 0.156$\pm$0.008 &   0.0$-$ 10.0 & 0.00726$\pm$0.00013
\\ 0.05$-$0.10 & 0.171$\pm$0.009 &  10.0$-$ 20.0 & 0.00780$\pm$0.00013
\\ 0.10$-$0.15 & 0.177$\pm$0.009 &  20.0$-$ 30.0 & 0.00780$\pm$0.00013
\\ 0.15$-$0.20 & 0.181$\pm$0.009 &  30.0$-$ 40.0 & 0.00673$\pm$0.00012
\\ 0.20$-$0.25 & 0.199$\pm$0.009 &  40.0$-$ 50.0 & 0.00561$\pm$0.00011
\\ 0.25$-$0.30 & 0.196$\pm$0.009 &  50.0$-$ 60.0 & 0.00440$\pm$0.00010
\\ 0.30$-$0.35 & 0.252$\pm$0.010 &  60.0$-$ 70.0 & 0.00376$\pm$0.00009
\\ 0.35$-$0.40 & 0.262$\pm$0.011 &  70.0$-$ 80.0 & 0.00327$\pm$0.00008
\\ 0.40$-$0.45 & 0.291$\pm$0.011 &  80.0$-$ 90.0 & 0.00325$\pm$0.00008
\\ 0.45$-$0.50 & 0.359$\pm$0.012 &  90.0$-$100.0 & 0.00317$\pm$0.00008
\\ 0.50$-$0.55 & 0.416$\pm$0.013 & 100.0$-$110.0 & 0.00336$\pm$0.00009
\\ 0.55$-$0.60 & 0.537$\pm$0.015 & 110.0$-$120.0 & 0.00384$\pm$0.00009
\\ 0.60$-$0.65 & 0.677$\pm$0.017 & 120.0$-$130.0 & 0.00444$\pm$0.00010
\\ 0.65$-$0.70 & 0.928$\pm$0.020 & 130.0$-$140.0 & 0.00553$\pm$0.00011
\\ 0.70$-$0.75 & 1.212$\pm$0.023 & 140.0$-$150.0 & 0.00686$\pm$0.00012
\\ 0.75$-$0.80 & 1.692$\pm$0.027 & 150.0$-$160.0 & 0.00792$\pm$0.00013
\\ 0.80$-$0.85 & 2.160$\pm$0.031 & 160.0$-$170.0 & 0.00776$\pm$0.00013
\\ 0.85$-$0.90 & 2.767$\pm$0.035 & 170.0$-$180.0 & 0.00722$\pm$0.00012
\\ 0.90$-$0.95 & 3.636$\pm$0.040 & &
\\ 0.95$-$1.00 & 3.729$\pm$0.040 & &
\\
     \hline\end{tabular}
    \caption{The measured $\cos\theta^*_3$ and $\psi^*$ distributions with
             their statistical errors
             for the three--jet events. The estimated systematic
             uncertainty is 6\%.}
   \label{tab:3jangle}
   \end{center}
 \end{table}

 \begin{table}[htb]
   \begin{center}
      \begin{tabular}{|c|c||c|c||c|c|} \hline
   $\mu_{34}$  & $1/N dN/d\mu_{34}$ & $\mu_{35}$
   & $1/N dN/d\mu_{35}$ & $\mu_{45}$  & $1/N dN/d\mu_{45}$ \\ \hline
0.58$-$0.60 & 0.284$\pm$0.018 & 0.22$-$0.24 & 0.099$\pm$0.010
& 0.08$-$0.10 & 0.090$\pm$0.010 \\ 0.60$-$0.62 & 0.815$\pm$0.030
& 0.24$-$0.26 & 0.204$\pm$0.015 & 0.10$-$0.12 & 0.265$\pm$0.017
\\ 0.62$-$0.64 & 1.360$\pm$0.038 & 0.26$-$0.28 & 0.336$\pm$0.019
& 0.12$-$0.14 & 0.493$\pm$0.023 \\ 0.64$-$0.66 & 2.055$\pm$0.047
& 0.28$-$0.30 & 0.571$\pm$0.025 & 0.14$-$0.16 & 0.843$\pm$0.030
\\ 0.66$-$0.68 & 2.626$\pm$0.053 & 0.30$-$0.32 & 0.843$\pm$0.030
& 0.16$-$0.18 & 1.296$\pm$0.037 \\ 0.68$-$0.70 & 3.535$\pm$0.062
& 0.32$-$0.34 & 1.271$\pm$0.037 & 0.18$-$0.20 & 1.683$\pm$0.043
\\ 0.70$-$0.72 & 3.719$\pm$0.063 & 0.34$-$0.36 & 1.647$\pm$0.042
& 0.20$-$0.22 & 2.215$\pm$0.049 \\ 0.72$-$0.74 & 3.629$\pm$0.063
& 0.36$-$0.38 & 1.905$\pm$0.045 & 0.22$-$0.24 & 2.666$\pm$0.054
\\ 0.74$-$0.76 & 3.494$\pm$0.061 & 0.38$-$0.40 & 2.170$\pm$0.048
& 0.24$-$0.26 & 3.140$\pm$0.058 \\ 0.76$-$0.78 & 3.497$\pm$0.061
& 0.40$-$0.42 & 2.340$\pm$0.050 & 0.26$-$0.28 & 3.317$\pm$0.060
\\ 0.78$-$0.80 & 3.491$\pm$0.061 & 0.42$-$0.44 & 2.465$\pm$0.052
& 0.28$-$0.30 & 3.684$\pm$0.063 \\ 0.80$-$0.82 & 3.514$\pm$0.062
& 0.44$-$0.46 & 2.628$\pm$0.053 & 0.30$-$0.32 & 3.716$\pm$0.063
\\ 0.82$-$0.84 & 3.518$\pm$0.062 & 0.46$-$0.48 & 2.777$\pm$0.055
& 0.32$-$0.34 & 3.592$\pm$0.062 \\ 0.84$-$0.86 & 3.625$\pm$0.063
& 0.48$-$0.50 & 2.914$\pm$0.056 & 0.34$-$0.36 & 3.434$\pm$0.061
\\ 0.86$-$0.88 & 3.467$\pm$0.061 & 0.50$-$0.52 & 3.297$\pm$0.060
& 0.36$-$0.38 & 3.293$\pm$0.060 \\ 0.88$-$0.90 & 3.273$\pm$0.059
& 0.52$-$0.54 & 3.332$\pm$0.060 & 0.38$-$0.40 & 2.916$\pm$0.056
\\ 0.90$-$0.92 & 2.296$\pm$0.050 & 0.54$-$0.56 & 3.750$\pm$0.064
& 0.40$-$0.42 & 2.582$\pm$0.053 \\ 0.92$-$0.94 & 1.267$\pm$0.037
& 0.56$-$0.58 & 4.024$\pm$0.066 & 0.42$-$0.44 & 2.369$\pm$0.051
\\ 0.94$-$0.96 & 0.482$\pm$0.023 & 0.58$-$0.60 & 3.919$\pm$0.065
& 0.44$-$0.46 & 2.102$\pm$0.048 \\ &                        &
0.60$-$0.62 & 3.390$\pm$0.061 & 0.46$-$0.48 & 1.887$\pm$0.045
\\ &                        & 0.62$-$0.64 & 2.679$\pm$0.054            &
0.48$-$0.50 & 1.584$\pm$0.041 \\ &                        &
0.64$-$0.66 & 1.945$\pm$0.046 & 0.50$-$0.52 & 1.176$\pm$0.036
\\ &                        & 0.66$-$0.68 & 1.136$\pm$0.035            &
0.52$-$0.54 & 0.907$\pm$0.031 \\ &                        &
0.68$-$0.70 & 0.301$\pm$0.018 & 0.54$-$0.56 & 0.565$\pm$0.025
\\ &                        & &                       &
0.56$-$0.58 & 0.154$\pm$0.013 \\
     \hline\end{tabular}
    \caption{The measured $\mu_{34}$, $\mu_{35}$ and
             $\mu_{45}$ distributions with their statistical errors
             for the three--jet events. The estimated systematic uncertainty
             is 6\%.}
   \label{tab:3jmass}

\vspace*{1.5cm}
      \begin{tabular}{|c|c||c|c||c|c||c|c|} \hline
  $x_3$  &  $1/N dN/dx_3$ &  $x_4$  &  $1/N dN/dx_4$ &  $x_5$
         &  $1/N dN/dx_5$ &  $x_6$  &  $1/N dN/dx_6$ \\ \hline
0.525$-$0.550 & 0.14$\pm$0.03 & 0.39$-$0.42 & 0.13$\pm$0.02   & 0.100$-$0.125 &
0.15$\pm$0.03 & 0.050$-$0.075 & 0.16$\pm$0.03\\ 0.550$-$0.575 & 0.43$\pm$0.05 &
0.42$-$0.45 & 0.36$\pm$0.04   & 0.125$-$0.150 & 0.30$\pm$0.04 & 0.075$-$0.100 &
0.97$\pm$0.07\\ 0.575$-$0.600 & 1.12$\pm$0.07 & 0.45$-$0.48 & 1.12$\pm$0.07   &
0.150$-$0.175 & 0.46$\pm$0.05 & 0.100$-$0.125 & 1.90$\pm$0.10\\ 0.600$-$0.625 &
1.51$\pm$0.09 & 0.48$-$0.51 & 2.35$\pm$0.10   & 0.175$-$0.200 & 0.81$\pm$0.06 &
0.125$-$0.150 & 3.08$\pm$0.12\\ 0.625$-$0.650 & 2.18$\pm$0.10 & 0.51$-$0.54 &
3.65$\pm$0.12   & 0.200$-$0.225 & 1.10$\pm$0.07 & 0.150$-$0.175 &
3.82$\pm$0.14\\ 0.650$-$0.675 & 2.77$\pm$0.12 & 0.54$-$0.57 & 4.49$\pm$0.14   &
0.225$-$0.250 & 1.33$\pm$0.08 & 0.175$-$0.200 & 4.49$\pm$0.15\\ 0.675$-$0.700 &
2.91$\pm$0.12 & 0.57$-$0.60 & 4.57$\pm$0.14   & 0.250$-$0.275 & 1.77$\pm$0.09 &
0.200$-$0.225 & 4.48$\pm$0.15\\ 0.700$-$0.725 & 3.39$\pm$0.13 & 0.60$-$0.63 &
3.94$\pm$0.13   & 0.275$-$0.300 & 2.03$\pm$0.10 & 0.225$-$0.250 &
4.25$\pm$0.15\\ 0.725$-$0.750 & 3.40$\pm$0.13 & 0.63$-$0.66 & 3.25$\pm$0.12   &
0.300$-$0.325 & 2.51$\pm$0.11 & 0.250$-$0.275 & 3.69$\pm$0.14\\ 0.750$-$0.775 &
3.53$\pm$0.13 & 0.66$-$0.69 & 2.67$\pm$0.10   & 0.325$-$0.350 & 2.82$\pm$0.12 &
0.275$-$0.300 & 3.26$\pm$0.13\\ 0.775$-$0.800 & 3.65$\pm$0.13 & 0.69$-$0.72 &
2.09$\pm$0.09   & 0.350$-$0.375 & 3.10$\pm$0.12 & 0.300$-$0.325 &
2.95$\pm$0.12\\ 0.800$-$0.825 & 3.52$\pm$0.13 & 0.72$-$0.75 & 1.68$\pm$0.08   &
0.375$-$0.400 & 3.69$\pm$0.14 & 0.325$-$0.350 & 2.25$\pm$0.11\\ 0.825$-$0.850 &
3.31$\pm$0.13 & 0.75$-$0.78 & 1.24$\pm$0.07   & 0.400$-$0.425 & 4.22$\pm$0.14 &
0.350$-$0.375 & 1.82$\pm$0.10\\ 0.850$-$0.875 & 2.83$\pm$0.12 & 0.78$-$0.81 &
0.89$\pm$0.06   & 0.425$-$0.450 & 4.33$\pm$0.15 & 0.375$-$0.400 &
1.35$\pm$0.08\\ 0.875$-$0.900 & 2.35$\pm$0.11 & 0.81$-$0.84 & 0.54$\pm$0.05   &
0.450$-$0.475 & 4.08$\pm$0.14 & 0.400$-$0.425 & 0.83$\pm$0.06\\ 0.900$-$0.925 &
1.59$\pm$0.09 & 0.84$-$0.87 & 0.23$\pm$0.03   & 0.475$-$0.500 & 3.39$\pm$0.13 &
0.425$-$0.450 & 0.52$\pm$0.05\\ 0.925$-$0.950 & 1.03$\pm$0.07 & 0.87$-$0.90 &
0.12$\pm$0.02   & 0.500$-$0.525 & 2.22$\pm$0.10 & 0.450$-$0.475 &
0.15$\pm$0.03\\ 0.950$-$0.975 & 0.30$\pm$0.04 & &                  &
0.525$-$0.550 & 1.08$\pm$0.07 &            & \\ &
&            &                  & 0.550$-$0.575 & 0.52$\pm$0.05 & &
\\ &                  &            &                  & 0.575$-$0.600 &
0.11$\pm$0.02 &            &                 \\
      \hline\end{tabular}
    \caption{The measured jet energy fraction distribitions for the
             four--jet events in their center--of--mass system.
             Errors are statistical only. The estimated systematic
             uncertainty is 6\%.}
   \label{tab:4jxvar}
   \end{center}
 \end{table}

 \begin{table}[htb]
   \begin{center}
      \begin{tabular}{|c||c|c|c|c|} \hline
  $\cos\theta^*_i$ & $1/N dN/d\cos\theta^*_3$
           & $1/N dN/d\cos\theta^*_4$
           & $1/N dN/d\cos\theta^*_5$
           & $1/N dN/d\cos\theta^*_6$ \\ \hline
0.00$-$0.05 &  0.18$\pm$0.02 &  0.27$\pm$0.03 &  0.60$\pm$0.04 &
0.79$\pm$0.04\\ 0.05$-$0.10 &  0.17$\pm$0.02 &  0.24$\pm$0.02 &  0.56$\pm$0.04
&  0.90$\pm$0.05\\ 0.10$-$0.15 &  0.13$\pm$0.02 &  0.29$\pm$0.03 &
0.56$\pm$0.04 &  1.05$\pm$0.05\\ 0.15$-$0.20 &  0.23$\pm$0.02 &  0.27$\pm$0.03
&  0.60$\pm$0.04 &  0.93$\pm$0.05\\ 0.20$-$0.25 &  0.18$\pm$0.02 &
0.27$\pm$0.03 &  0.61$\pm$0.04 &  0.95$\pm$0.05\\ 0.25$-$0.30 &  0.23$\pm$0.02
&  0.34$\pm$0.03 &  0.58$\pm$0.04 &  0.94$\pm$0.05\\ 0.30$-$0.35 &
0.22$\pm$0.02 &  0.33$\pm$0.03 &  0.74$\pm$0.04 &  0.91$\pm$0.05\\ 0.35$-$0.40
&  0.28$\pm$0.03 &  0.30$\pm$0.03 &  0.67$\pm$0.04 &  0.93$\pm$0.05\\
0.40$-$0.45 &  0.35$\pm$0.03 &  0.43$\pm$0.03 &  0.74$\pm$0.04 &
0.95$\pm$0.05\\ 0.45$-$0.50 &  0.35$\pm$0.03 &  0.46$\pm$0.03 &  0.72$\pm$0.04
&  0.94$\pm$0.05\\ 0.50$-$0.55 &  0.38$\pm$0.03 &  0.47$\pm$0.03 &
0.85$\pm$0.05 &  1.06$\pm$0.05\\ 0.55$-$0.60 &  0.49$\pm$0.03 &  0.58$\pm$0.04
&  0.74$\pm$0.04 &  1.16$\pm$0.05\\ 0.60$-$0.65 &  0.63$\pm$0.04 &
0.64$\pm$0.04 &  0.94$\pm$0.05 &  1.13$\pm$0.05\\ 0.65$-$0.70 &  0.79$\pm$0.04
&  0.78$\pm$0.04 &  0.94$\pm$0.05 &  1.18$\pm$0.05\\ 0.70$-$0.75 &
0.93$\pm$0.05 &  0.88$\pm$0.05 &  1.11$\pm$0.05 &  1.17$\pm$0.05\\ 0.75$-$0.80
&  1.21$\pm$0.05 &  1.06$\pm$0.05 &  1.30$\pm$0.06 &  1.19$\pm$0.05\\
0.80$-$0.85 &  1.68$\pm$0.06 &  1.47$\pm$0.06 &  1.62$\pm$0.06 &
1.28$\pm$0.06\\ 0.85$-$0.90 &  2.32$\pm$0.08 &  2.09$\pm$0.07 &  2.19$\pm$0.07
&  1.24$\pm$0.06\\ 0.90$-$0.95 &  3.62$\pm$0.09 &  3.53$\pm$0.09 &
2.53$\pm$0.08 &  0.98$\pm$0.05\\ 0.95$-$1.00 &  5.63$\pm$0.12 &  5.29$\pm$0.11
&  1.39$\pm$0.06 &  0.32$\pm$0.03\\
     \hline\end{tabular}
    \caption{The measured jet cosine distributions for the four--jet
             events in their center--of--mass system.
             Errors are statistical only. The estimated systematic
             uncertainty is 6\%.}
   \label{tab:4jcos}

\vspace*{1.5cm}
     \begin{tabular}{|c||c|c|c|c|c|c|} \hline
  $\cos\omega_{ij}$ & $1/N dN/d\cos\omega_{34}$
           & $1/N dN/d\cos\omega_{35}$
           & $1/N dN/d\cos\omega_{36}$
           & $1/N dN/d\cos\omega_{45}$
           & $1/N dN/d\cos\omega_{46}$
           & $1/N dN/d\cos\omega_{56}$ \\ \hline
-1.0$-$ -0.9 & 4.557$\pm$0.075 & 1.180$\pm$0.038 & 0.550$\pm$0.026 &
0.043$\pm$0.007 & 0.407$\pm$0.022 & 0.782$\pm$0.031\\ -0.9$-$ -0.8 &
2.245$\pm$0.053 & 1.475$\pm$0.043 & 0.483$\pm$0.024 & 0.151$\pm$0.014 &
0.505$\pm$0.025 & 0.539$\pm$0.026\\ -0.8$-$ -0.7 & 1.211$\pm$0.039 &
1.409$\pm$0.042 & 0.515$\pm$0.025 & 0.256$\pm$0.018 & 0.531$\pm$0.026 &
0.624$\pm$0.028\\ -0.7$-$ -0.6 & 0.815$\pm$0.032 & 1.150$\pm$0.038 &
0.577$\pm$0.027 & 0.300$\pm$0.019 & 0.535$\pm$0.026 & 0.596$\pm$0.027\\ -0.6$-$
-0.5 & 0.451$\pm$0.024 & 0.958$\pm$0.034 & 0.651$\pm$0.028 & 0.334$\pm$0.020 &
0.542$\pm$0.026 & 0.545$\pm$0.026\\ -0.5$-$ -0.4 & 0.290$\pm$0.019 &
0.833$\pm$0.032 & 0.612$\pm$0.028 & 0.411$\pm$0.023 & 0.546$\pm$0.026 &
0.594$\pm$0.027\\ -0.4$-$ -0.3 & 0.197$\pm$0.016 & 0.774$\pm$0.031 &
0.651$\pm$0.028 & 0.422$\pm$0.023 & 0.566$\pm$0.026 & 0.603$\pm$0.027\\ -0.3$-$
-0.2 & 0.097$\pm$0.011 & 0.555$\pm$0.026 & 0.664$\pm$0.029 & 0.511$\pm$0.025 &
0.531$\pm$0.026 & 0.629$\pm$0.028\\ -0.2$-$ -0.1 & 0.058$\pm$0.008 &
0.448$\pm$0.024 & 0.688$\pm$0.029 & 0.573$\pm$0.027 & 0.550$\pm$0.026 &
0.662$\pm$0.029\\ -0.1$-$  0.0 & 0.026$\pm$0.006 & 0.391$\pm$0.022 &
0.643$\pm$0.028 & 0.619$\pm$0.028 & 0.620$\pm$0.028 & 0.678$\pm$0.029\\ 0.0$-$
0.1 & 0.021$\pm$0.005 & 0.264$\pm$0.018 & 0.701$\pm$0.029 & 0.719$\pm$0.030 &
0.600$\pm$0.027 & 0.737$\pm$0.030\\ 0.1$-$  0.2 & 0.022$\pm$0.005 &
0.208$\pm$0.016 & 0.697$\pm$0.029 & 0.745$\pm$0.030 & 0.636$\pm$0.028 &
0.738$\pm$0.030\\ 0.2$-$  0.3 & 0.007$\pm$0.003 & 0.129$\pm$0.013 &
0.588$\pm$0.027 & 0.774$\pm$0.031 & 0.603$\pm$0.027 & 0.701$\pm$0.029\\ 0.3$-$
0.4 & 0.001$\pm$0.001 & 0.100$\pm$0.011 & 0.511$\pm$0.025 & 0.812$\pm$0.032 &
0.597$\pm$0.027 & 0.459$\pm$0.024\\ 0.4$-$  0.5 &                 &
0.062$\pm$0.009 & 0.404$\pm$0.022 & 0.747$\pm$0.030 & 0.573$\pm$0.027 &
0.396$\pm$0.022\\ 0.5$-$  0.6 &                 & 0.037$\pm$0.007 &
0.360$\pm$0.021 & 0.745$\pm$0.030 & 0.513$\pm$0.025 & 0.311$\pm$0.020\\ 0.6$-$
0.7 &                 & 0.020$\pm$0.005 & 0.307$\pm$0.019 & 0.672$\pm$0.029 &
0.426$\pm$0.023 & 0.222$\pm$0.017\\ 0.7$-$  0.8 &                 &
0.005$\pm$0.002 & 0.224$\pm$0.017 & 0.626$\pm$0.028 & 0.397$\pm$0.022 &
0.119$\pm$0.012\\ 0.8$-$  0.9 &                 & 0.002$\pm$0.002 &
0.144$\pm$0.013 & 0.420$\pm$0.023 & 0.262$\pm$0.018 & 0.054$\pm$0.008\\ 0.9$-$
1.0 &                 &                 & 0.031$\pm$0.006 & 0.118$\pm$0.012 &
0.058$\pm$0.008 & 0.011$\pm$0.004\\
     \hline\end{tabular}
    \caption{The measured distribution of the cosine of space angles
     between pairs of jets for the four--jet events in their
     center--of--mass system. Errors are statistical only.
     The estimated systematic uncertainty is 6\%.}
   \label{tab:4jangle}
   \end{center}
 \end{table}

 \begin{table}[htb]
   \begin{center}
    \begin{tabular}{|c||c|c|c|c|c|c|} \hline
  $\mu_{ij}$ & $1/N dN/d\mu_{34}$
           & $1/N dN/d\mu_{35}$
           & $1/N dN/d\mu_{36}$
           & $1/N dN/d\mu_{45}$
           & $1/N dN/d\mu_{46}$
           & $1/N dN/d\mu_{56}$ \\ \hline
0.00$-$0.03 &                &                &                &
&                &               \\ 0.03$-$0.06 &                &
&                &  0.02$\pm$0.01 &  0.01$\pm$0.01 &  0.02$\pm$0.01\\
0.06$-$0.09 &                &                &  0.11$\pm$0.02 &  0.24$\pm$0.03
&  0.31$\pm$0.04 &  0.55$\pm$0.05\\ 0.09$-$0.12 &                &
&  0.45$\pm$0.04 &  0.77$\pm$0.06 &  0.99$\pm$0.06 &  1.76$\pm$0.09\\
0.12$-$0.15 &                &  0.01$\pm$0.01 &  1.11$\pm$0.07 &  1.49$\pm$0.08
&  1.95$\pm$0.09 &  3.30$\pm$0.12\\ 0.15$-$0.18 &                &
0.05$\pm$0.01 &  1.58$\pm$0.08 &  2.13$\pm$0.09 &  2.79$\pm$0.11 &
4.52$\pm$0.14\\ 0.18$-$0.21 &                &  0.09$\pm$0.02 &  2.06$\pm$0.09
&  2.51$\pm$0.10 &  3.44$\pm$0.12 &  5.25$\pm$0.15\\ 0.21$-$0.24 &
&  0.19$\pm$0.03 &  3.03$\pm$0.11 &  2.77$\pm$0.11 &  4.28$\pm$0.13 &
4.80$\pm$0.14\\ 0.24$-$0.27 &                &  0.33$\pm$0.04 &  3.38$\pm$0.12
&  3.00$\pm$0.11 &  4.33$\pm$0.13 &  3.99$\pm$0.13\\ 0.27$-$0.30 &
&  0.62$\pm$0.05 &  3.62$\pm$0.12 &  3.47$\pm$0.12 &  3.99$\pm$0.13 &
3.14$\pm$0.11\\ 0.30$-$0.33 &  0.01$\pm$0.01 &  1.23$\pm$0.07 &  3.90$\pm$0.13
&  3.36$\pm$0.12 &  3.61$\pm$0.12 &  2.32$\pm$0.10\\ 0.33$-$0.36 &
0.06$\pm$0.02 &  1.57$\pm$0.08 &  3.49$\pm$0.12 &  2.94$\pm$0.11 &
2.88$\pm$0.11 &  1.65$\pm$0.08\\ 0.36$-$0.39 &  0.09$\pm$0.02 &  2.14$\pm$0.09
&  3.07$\pm$0.11 &  2.99$\pm$0.11 &  2.13$\pm$0.09 &  1.05$\pm$0.07\\
0.39$-$0.42 &  0.12$\pm$0.02 &  2.91$\pm$0.11 &  2.82$\pm$0.11 &  2.72$\pm$0.11
&  1.40$\pm$0.08 &  0.60$\pm$0.05\\ 0.42$-$0.45 &  0.25$\pm$0.03 &
3.53$\pm$0.12 &  2.03$\pm$0.09 &  2.26$\pm$0.10 &  0.84$\pm$0.06 &
0.29$\pm$0.03\\ 0.45$-$0.48 &  0.61$\pm$0.05 &  4.35$\pm$0.13 &  1.59$\pm$0.08
&  1.47$\pm$0.08 &  0.37$\pm$0.04 &  0.09$\pm$0.02\\ 0.48$-$0.51 &
1.06$\pm$0.07 &  4.92$\pm$0.14 &  0.81$\pm$0.06 &  0.95$\pm$0.06 &
0.02$\pm$0.01 &               \\ 0.51$-$0.54 &  1.64$\pm$0.08 &  4.60$\pm$0.14
&  0.25$\pm$0.03 &  0.26$\pm$0.03 &                &               \\
0.54$-$0.57 &  2.74$\pm$0.11 &  3.63$\pm$0.12 &  0.04$\pm$0.01 &  0.01$\pm$0.01
&                &               \\ 0.57$-$0.60 &  3.86$\pm$0.13 &
2.03$\pm$0.09 &                &                &                &
\\ 0.60$-$0.63 &  3.80$\pm$0.13 &  0.88$\pm$0.06 &                &
&                &               \\ 0.63$-$0.66 &  4.15$\pm$0.13 &
0.24$\pm$0.03 &                &                &                &
\\ 0.66$-$0.69 &  3.68$\pm$0.12 &  0.01$\pm$0.01 &                &
&                &               \\ 0.69$-$0.72 &  3.15$\pm$0.11 &
&                &                &                &               \\
0.72$-$0.75 &  2.65$\pm$0.10 &                &                &
&                &               \\ 0.75$-$0.78 &  2.05$\pm$0.09 &
&                &                &                &               \\
0.78$-$0.81 &  1.57$\pm$0.08 &                &                &
&                &               \\ 0.81$-$0.84 &  0.98$\pm$0.06 &
&                &                &                &               \\
0.84$-$0.87 &  0.56$\pm$0.05 &                &                &
&                &               \\ 0.87$-$0.90 &  0.30$\pm$0.04 &
&                &                &                &               \\
     \hline\end{tabular}
    \caption{The measured distribution of scaled jet pair masses for
             the four--jet events in their center--of--mass system.
             Errors are statistical only. The estimated systematic
             uncertainty is 6\%.}
   \label{tab:4jmass}

\vspace*{1.5cm}
      \begin{tabular}{|c|c||c|c|}\hline
   $\chi_{BZ}$  & $1/N dN/d\chi_{BZ}$ & $\cos\theta_{NR}$  &
                  $1/N dN/d\cos\theta_{NR}$ \\ \hline
0.0$-$10.0 & 0.0073$\pm$0.0003        &   0.0$-$0.1 & 1.020$\pm$0.036
\\ 10.0$-$20.0 & 0.0079$\pm$0.0003 &   0.1$-$0.2 & 1.068$\pm$0.036
\\ 20.0$-$30.0 & 0.0089$\pm$0.0003 &   0.2$-$0.3 & 0.942$\pm$0.034
\\ 30.0$-$40.0 & 0.0102$\pm$0.0004 &   0.3$-$0.4 & 0.982$\pm$0.035
\\ 40.0$-$50.0 & 0.0107$\pm$0.0004 &   0.4$-$0.5 & 0.993$\pm$0.035
\\ 50.0$-$60.0 & 0.0124$\pm$0.0004 &   0.5$-$0.6 & 0.988$\pm$0.035
\\ 60.0$-$70.0 & 0.0132$\pm$0.0004 &   0.6$-$0.7 & 0.956$\pm$0.034
\\ 70.0$-$80.0 & 0.0146$\pm$0.0004 &   0.7$-$0.8 & 0.986$\pm$0.035
\\ 80.0$-$90.0 & 0.0148$\pm$0.0004 &   0.8$-$0.9 & 0.993$\pm$0.035
\\ &                          &   0.9$-$1.0 & 1.072$\pm$0.036         \\
     \hline\end{tabular}
    \caption{The measured $\chi_{BZ}$ and $\cos\theta_{NR}$ distributions
             with their statistical errors
             for the four--jet events. The estimated systematic
             uncertainty is 6\%.}
   \label{tab:4jbznr}
   \end{center}
 \end{table}

\end{document}